\documentclass[12pt]{article}
\usepackage{epstopdf}
\usepackage{epsfig}
\usepackage{amsmath}
\usepackage{hhline}
\usepackage{amssymb}
\usepackage{times}
\usepackage{cite}
\usepackage[]{lineno}
\usepackage{multirow}
\usepackage{verbatim} 
\usepackage{url} 
\usepackage{captcont} 

\newlength{\dinwidth}
\newlength{\dinmargin}
\begin{document}  
\newcommand{\pom}{{I\!\!P}}
\newcommand{\reg}{{I\!\!R}}
\newcommand{\slowpi}{\pi_{\mathit{slow}}}
\newcommand{\fiidiii}{F_2^{D(3)}}
\newcommand{\fiidiiiarg}{\fiidiii\,(\beta,\,Q^2,\,x)}
\newcommand{\n}{1.19\pm 0.06 (stat.) \pm0.07 (syst.)}
\newcommand{\nz}{1.30\pm 0.08 (stat.)^{+0.08}_{-0.14} (syst.)}
\newcommand{\fiidiiiful}{F_2^{D(4)}\,(\beta,\,Q^2,\,x,\,t)}
\newcommand{\fiipom}{\tilde F_2^D}
\newcommand{\ALPHA}{1.10\pm0.03 (stat.) \pm0.04 (syst.)}
\newcommand{\ALPHAZ}{1.15\pm0.04 (stat.)^{+0.04}_{-0.07} (syst.)}
\newcommand{\fiipomarg}{\fiipom\,(\beta,\,Q^2)}
\newcommand{\pomflux}{f_{\pom / p}}
\newcommand{\nxpom}{1.19\pm 0.06 (stat.) \pm0.07 (syst.)}
\newcommand {\gapprox}
   {\raisebox{-0.7ex}{$\stackrel {\textstyle>}{\sim}$}}
\newcommand {\lapprox}
   {\raisebox{-0.7ex}{$\stackrel {\textstyle<}{\sim}$}}
\def\gsim{\,\lower.25ex\hbox{$\scriptstyle\sim$}\kern-1.30ex%
\raise 0.55ex\hbox{$\scriptstyle >$}\,}
\def\lsim{\,\lower.25ex\hbox{$\scriptstyle\sim$}\kern-1.30ex%
\raise 0.55ex\hbox{$\scriptstyle <$}\,}
\newcommand{\pomfluxarg}{f_{\pom / p}\,(x_\pom)}
\newcommand{\dsf}{\mbox{$F_2^{D(3)}$}}
\newcommand{\dsfva}{\mbox{$F_2^{D(3)}(\beta,Q^2,x_{I\!\!P})$}}
\newcommand{\dsfvb}{\mbox{$F_2^{D(3)}(\beta,Q^2,x)$}}
\newcommand{\dsfpom}{$F_2^{I\!\!P}$}
\newcommand{\gap}{\stackrel{>}{\sim}}
\newcommand{\lap}{\stackrel{<}{\sim}}
\newcommand{\fem}{$F_2^{em}$}
\newcommand{\tsnmp}{$\tilde{\sigma}_{NC}(e^{\mp})$}
\newcommand{\tsnm}{$\tilde{\sigma}_{NC}(e^-)$}
\newcommand{\tsnp}{$\tilde{\sigma}_{NC}(e^+)$}
\newcommand{\st}{$\star$}
\newcommand{\sst}{$\star \star$}
\newcommand{\ssst}{$\star \star \star$}
\newcommand{\sssst}{$\star \star \star \star$}
\newcommand{\tw}{\theta_W}
\newcommand{\sw}{\sin{\theta_W}}
\newcommand{\cw}{\cos{\theta_W}}
\newcommand{\sww}{\sin^2{\theta_W}}
\newcommand{\cww}{\cos^2{\theta_W}}
\newcommand{\trm}{m_{\perp}}
\newcommand{\trp}{p_{\perp}}
\newcommand{\trmm}{m_{\perp}^2}
\newcommand{\trpp}{p_{\perp}^2}
\newcommand{\alp}{\alpha_s}

\newcommand{\alps}{\alpha_s}
\newcommand{\sqrts}{$\sqrt{s}$}
\newcommand{\LO}{$O(\alpha_s^0)$}
\newcommand{\Oa}{$O(\alpha_s)$}
\newcommand{\Oaa}{$O(\alpha_s^2)$}
\newcommand{\PT}{p_{\perp}}
\newcommand{\JPSI}{J/\psi}
\newcommand{\sh}{\hat{s}}
\newcommand{\uh}{\hat{u}}
\newcommand{\MP}{m_{J/\psi}}
\newcommand{\PO}{I\!\!P}
\newcommand{\xbj}{x}
\newcommand{\xpom}{x_{\PO}}
\newcommand{\ttbs}{\char'134}
\newcommand{\xpomlo}{3\times10^{-4}}  
\newcommand{\xpomup}{0.05}  
\newcommand{\dgr}{^\circ}
\newcommand{\pbarnt}{\,\mbox{{\rm pb$^{-1}$}}}
\newcommand{\gev}{\,\mbox{GeV}}
\newcommand{\WBoson}{\mbox{$W$}}
\newcommand{\fbarn}{\,\mbox{{\rm fb}}}
\newcommand{\fbarnt}{\,\mbox{{\rm fb$^{-1}$}}}
\newcommand{\dsdx}[1]{$d\sigma\!/\!d #1\,$}
\newcommand{\eV}{\mbox{e\hspace{-0.08em}V}}
%
%
\newcommand{\qsq}{\ensuremath{Q^2} }
\newcommand{\gevsq}{\ensuremath{\mathrm{GeV}^2} }
\newcommand{\et}{\ensuremath{E_t^*} }
\newcommand{\rap}{\ensuremath{\eta^*} }
\newcommand{\gp}{\ensuremath{\gamma^*}p }
\newcommand{\dsiget}{\ensuremath{{\rm d}\sigma_{ep}/{\rm d}E_t^*} }
\newcommand{\dsigrap}{\ensuremath{{\rm d}\sigma_{ep}/{\rm d}\eta^*} }

\newcommand{\dstar}{\ensuremath{D^*}}
\newcommand{\dstarp}{\ensuremath{D^{*+}}}
\newcommand{\dstarm}{\ensuremath{D^{*-}}}
\newcommand{\dstarpm}{\ensuremath{D^{*\pm}}}
\newcommand{\zDs}{\ensuremath{z(\dstar )}}
\newcommand{\Wgp}{\ensuremath{W_{\gamma p}}}
\newcommand{\ptds}{\ensuremath{p_t(\dstar )}}
\newcommand{\etads}{\ensuremath{\eta(\dstar )}}
\newcommand{\ptj}{\ensuremath{p_t(\mbox{jet})}}
\newcommand{\ptjn}[1]{\ensuremath{p_t(\mbox{jet$_{#1}$})}}
\newcommand{\etaj}{\ensuremath{\eta(\mbox{jet})}}
\newcommand{\detadsj}{\ensuremath{\eta(\dstar )\, \mbox{-}\, \etaj}}

\def\Journal#1#2#3#4{{#1} {\bf #2} (#3) #4}
\def\NCA{\em Nuovo Cimento}
\def\NIM{\em Nucl. Instrum. Methods}
\def\NIMA{{\em Nucl. Instrum. Methods} {\bf A}}
\def\NPB{{\em Nucl. Phys.}   {\bf B}}
\def\PLB{{\em Phys. Lett.}   {\bf B}}
\def\PRL{\em Phys. Rev. Lett.}
\def\PRD{{\em Phys. Rev.}    {\bf D}}
\def\ZPC{{\em Z. Phys.}      {\bf C}}
\def\EJC{{\em Eur. Phys. J.} {\bf C}}
\def\CPC{\em Comp. Phys. Commun.}

\begin{titlepage}

\noindent
\begin{flushleft}
{\tt DESY 13-012    \hfill    ISSN 0418-9833} \\
{\tt January 2013}                  \\
\end{flushleft}


\vspace{2cm}
\begin{center}
\begin{Large}

{\bf Measurement of Charged Particle Spectra in Deep-Inelastic \boldmath$ep$ Scattering at HERA}

\vspace{2cm}

H1 Collaboration

\end{Large}
\end{center}

\vspace{2cm}

\begin{abstract}

Charged particle production  in deep-inelastic $ep$ scattering is measured with the H1 detector at HERA. The kinematic range of the analysis covers low photon virtualities, \mbox{$5 < Q^2 < 100$ GeV$^2$}, and small values of Bjorken-$x$, $10^{-4} < x < 10^{-2}$. The analysis is performed in the hadronic centre-of-mass system. The charged particle densities are measured as a function of pseudorapidity ($\eta^*$) and transverse momentum ($p_T^*$) in the range 
$0<\eta^*<5$ and $0<p_T^* < 10$ GeV differentially in $x$ and $Q^2$. The data are compared to predictions from different Monte Carlo generators implementing various options for hadronisation and parton evolutions.

\end{abstract}

\vspace{1.5cm}

\begin{center}
Submitted to \EJC \;\;
\end{center}

\end{titlepage}

\begin{flushleft}

C.~Alexa$^{5}$,                
V.~Andreev$^{25}$,             
A.~Baghdasaryan$^{37}$,        
S.~Baghdasaryan$^{37}$,        
W.~Bartel$^{11}$,              
K.~Begzsuren$^{34}$,           
A.~Belousov$^{25}$,            
P.~Belov$^{11}$,               
V.~Boudry$^{28}$,              
I.~Bozovic-Jelisavcic$^{2}$
% ...


\bigskip{\it
% ...
$ ^{2}$ Vinca Institute of Nuclear Sciences, University of Belgrade,
          1100 Belgrade, Serbia \\
% ...
$ ^{5}$ National Institute for Physics and Nuclear Engineering (NIPNE) ,
          Bucharest, Romania$^{ k}$ \\
% ...
$ ^{11}$ DESY, Hamburg, Germany \\
% ...
$ ^{25}$ Lebedev Physical Institute, Moscow, Russia \\
% ...
$ ^{28}$ LLR, Ecole Polytechnique, CNRS/IN2P3, Palaiseau, France \\
% ...
$ ^{34}$ Institute of Physics and Technology of the Mongolian
          Academy of Sciences, Ulaanbaatar, Mongolia \\
% ...
$ ^{37}$ Yerevan Physics Institute, Yerevan, Armenia \\
% ...
$ ^k$ Supported by the Romanian National Authority for Scientific Research
      under the contract PN 09370101 \\
% ...
}\end{flushleft}

\newpage

\section{Introduction}

\begin{figure}[hhh]
\center
\epsfig{file=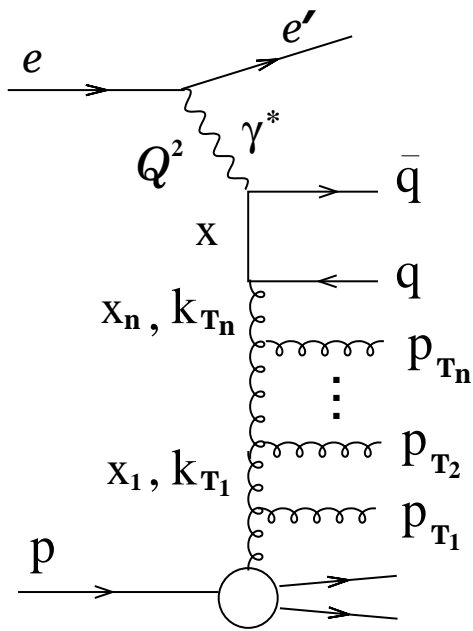,width=6.2cm,height=8.5cm}
\setlength{\unitlength}{1cm}
\caption{Generic diagram for deep-inelastic $ep$ scattering at small
  $x$. The transverse momenta of the emitted gluons are labeled as
  $p_{T,i}$, while the proton longitudinal
  momentum fractions and the transverse momenta carried
  by the propagating gluons are denoted by $x_i$ and
  $k_{T,i}$, respectively.}
\label{fig:ladder} 
\end{figure}

Deep-inelastic scattering (DIS) processes at the $ep$ collider HERA can access small values of Bjorken-$x$ at low four momentum transfers squared $Q^2$ of a few GeV$^2$.  
In the region of low $x$, characterised by high densities of gluons and sea quarks in the proton, the parton interaction with the virtual photon may originate from a cascade of partons emitted prior to the interaction as illustrated in \mbox{figure \ref{fig:ladder}}.
In perturbative Quantum Chromodynamics (QCD) such multi-parton emissions are described only within certain approximations valid in restricted phase space regions.
At sufficiently large $Q^2$ and not too small $x$ 
the Dokshitzer-Gribov-Lipatov-Altarelli-Parisi (DGLAP) \cite{DGLAP} evolution equation is expected to be a good approximation. The DGLAP equation corresponds to a strong ordering of the transverse momenta of the propagator partons, $k_{T,i}$ , with respect to the proton direction, which implies strong ordering of the transverse momenta of the emitted partons, $p_{T,i} \ll p_{T,i+1}$, in the parton cascade from the proton towards the virtual photon. At small $x$ the DGLAP approximation is expected to become inadequate and 
the Balitsky-Fadin-Kuraev-Lipatov (BFKL) \cite{BFKL} scheme may 
be more appropriate, which has no ordering in $k_{T}$ of the partons along the ladder. The Ciafaloni-Catani-Fiorani-Marchesini (CCFM) evolution \cite{CCFM} aims to unify the DGLAP and BFKL approaches. It introduces angular ordering of gluon emissions to implement coherence effects. At small $x$ the CCFM evolution equation is almost equivalent to the BFKL approach, while it reproduces the DGLAP equations for sufficiently large $x$ and $Q^2$.

Measurements of the proton structure function $F_2 (x, Q^2)$ \cite{F2} are well described by the Next-to-Leading-Order (NLO) or Next-to-Next-to-Leading Order (NNLO) DGLAP evolution \cite{F2-theory-MRST,F2-theory-CT,F2-theory-NNPDF,F2-theory-ABM}, suggesting that this observable may be too inclusive to exhibit signals for BFKL dynamics. Deviations from the $k_T$ ordering at HERA are observed in jet production \cite{dijet,forw-incl-jet-1}, transverse energy flow \cite{tr-energy-flow-1,tr-energy-flow-2}, forward jet production \cite{forw-jet-h1-zeus1,forw-jet-h1-zeus2,forw-jet-h1-zeus3} and measurements of forward $\pi^0$ production \cite{fwd-pi}.
Studies of the transverse momentum spectrum of charged particles have been proposed in \cite{Kuhlen-pt} as a more direct probe of the underlying parton dynamics. In that paper it has been shown with the help of QCD models that the high-$p_T$ tail is sensitive to parton radiation, while the contribution from hadronisation is small.
Previous measurements of the transverse momentum and pseudorapidity dependence of particle densities performed by the H1 collaboration \cite{pt-old}
were limited in statistical precision.

This paper presents a study of charged particle production in $ep$ collisions for 
\mbox{$5 < Q^2 < $} \mbox{$100$ GeV$^2$}. The analysis is performed in the hadronic centre-of-mass frame, i.e. in the virtual photon-proton rest frame. The charged particle densities as a function of pseudorapidity and transverse momentum are measured differentially in $x$ and $Q^2$.

The data set used for the analysis was collected with the H1 detector in 2006 when positrons and protons collided with energies of $27.6$ GeV and $920$ GeV, respectively, corresponding to a centre-of-mass energy of $\sqrt{s}=319$ GeV. The integrated luminosity of the data set is $88.6$ pb$^{-1}$, which is about seventy times larger than what was available for the previous H1 measurement \cite{pt-old}. This allows for a more detailed study of the dynamical features of parton evolution in the proton at small $x$.

\section{QCD models}
\label{sec-QCDmodels}

Parton cascade and hadronisation processes leading to charged particle production in $ep$ collisions are modeled using different Monte Carlo (MC) programs. 
Brief descriptions of the MC event generators considered in this analysis are given below.

\begin{itemize}
\item The {\sc Rapgap} \cite{RAPGAP} MC generator matches first order QCD matrix
  elements to DGLAP based leading logarithm approximations for parton showers with
  strongly ordered transverse momenta of subsequently emitted
  partons.  The factorisation and renormalisation scales are set to
  $\mu_f=\mu_r=\sqrt{Q^2+\hat{p_T}^2}$, where $\hat{p}_T$ is the transverse
  momentum of the outgoing hard parton from the matrix element
  in the centre-of-mass frame of the hard subsystem.

\item The {\sc Djangoh} \cite{DJANGOH} MC generator uses the Colour Dipole Model
  (CDM) as implemented in {\sc ARIADNE} \cite{ARIADNE}, which models first order QCD  processes and creates dipoles between coloured partons. Gluon emission is treated as
  radiation from these dipoles, and new dipoles are formed from the
  emitted gluons from which further radiation is possible. The
  radiation pattern of the dipoles includes interference effects, thus
  modelling gluon coherence. The transverse momenta of the emitted partons
  are not ordered, producing a configuration similar to
  the BFKL treatment of parton evolution \cite{CDM-BFKL}.

\item The {\sc Cascade} \cite{CASCADE} MC generator uses off-shell
  leading order QCD matrix elements, supplemented with parton
  emissions based on the CCFM evolution equation. The equation
  requires an unintegrated gluon density (see \cite{CASCADE}), which takes the
  transverse momenta of the propagators into account. 
  In contrast to the DGLAP evolution equation, the CCFM equation
  only contains gluon splitting $g \to g g$.

\item The Herwig++ \cite{herwig} MC program with the POWHEG (POsitive Weight Hardest  Emission) option \cite{POWHEG-DIS}, combines the full matrix element including
virtual corrections at $\mathcal{O}(\alpha_s)$ with a DGLAP-like parton shower simulation. The Herwig++ MC program uses the Coherent Parton Branching algorithm \cite{CoherentPB1,CoherentPB2} which is based on
colour coherence to suppress branchings outside an angular-ordered
region of phase space. Here, final state radiation is angular ordered
and initial state radiation is ordered in $E\cdot \theta \approx p_T$, where $E$
and $\theta$ are the energy and polar angle of the radiated parton, respectively.

\item Photoproduction background is generated with the {\sc Phojet} \cite{Phojet} program, which uses a two-component dual parton model \cite{Phojet1} including diffractive processes and vector meson production.

\end{itemize}

The {\sc Rapgap} and {\sc Djangoh} predictions are calculated using the CTEQ6L(LO) \cite{cteq6lo} set of Parton Distribution Functions (PDF), while {\sc Cascade} uses the default A0 unintegrated gluon density set \cite{A0}. The predictions of Herwig++ were obtained with the default PDF MRST 02 NLO \cite{MRST}. To simulate hadronisation the Lund string fragmentation model \cite{Lund} is used, as implemented in JETSET \cite{JETSETandPYTHIA} for {\sc Djangoh} and {\sc Pythia} \cite{PYTHIA} for both {\sc Rapgap} and {\sc Cascade}. 
The parameters of the Lund string fragmentation model used here are tuned to describe $e^+e^-$ results \cite{ALEPH}. The tuning was performed by the ALEPH collaboration using hadronic $Z$ decay data and the {\sc Pythia}6.1 simulation with Bose-Einstein correlations turned on. In addition, the tune obtained by the Professor tool \cite{profftune} using LEP data is also tested. Herwig++ incorporates the cluster model \cite{clustermodel1} of hadronisation, in which colour-singlet clusters of partons form after the perturbative phase and then decay into the observed hadrons.

{\sc Djangoh} and {\sc Rapgap} are also used together with the H1 detector simulation in order to determine the acceptance and efficiency and to estimate the systematic uncertainties associated with the measurement. The programs are interfaced to {\sc Heracles} \cite{HERACLES} to simulate the QED-radiative effects. The generated events are passed through a detailed simulation of the H1 detector response based on the {\sc Geant} simulation program \cite{GEANT} and are processed using the same reconstruction and analysis program chain as for data. For the determination of the detector effects both the {\sc Rapgap} and {\sc Djangoh} predictions are studied. Both models describe all relevant control distributions reasonably well \cite{myThesis}. To improve the determination of the detector corrections the transverse momentum and pseudorapidity of charged particles as well as inelasticity $y$, defined as $y=Q^2/(s\cdot x)$, are reweighted to the data \cite{myThesis}. The reweighting is applied to the generator quantities.

\section{Experimental method}

\subsection{H1 detector}
A full description of the H1 detector can be found elsewhere
\cite{H1det1,H1det2,SPACAL} and only the components most relevant for this
analysis are briefly mentioned here. The coordinate system of H1 is defined such that the positive $z$ axis is pointing in the direction of the proton beam (forward direction) and the nominal interaction point is located at $z = 0$. The polar angle $\theta$ is then defined with respect to this axis. The pseudorapidity is defined to be $\eta = - \ln (\tan(\theta/2))$.

Charged particles are measured within the central tracking detector
(CTD) in the polar angle range \mbox{$20^{\circ} < \theta < 165^{\circ} $}, which is also used to reconstruct the interaction vertex. The CTD comprises two large cylindrical jet chambers (CJCs), and the silicon vertex detector \cite{CST34,CST35}. The CTD is operated inside a $1.16$ T solenoidal magnetic field. The CJCs are separated by a cylindrical drift chamber which improves the $z$ coordinate reconstruction. A cylindrical multiwire proportional chamber \cite{CIP}, which is mainly used in the trigger, is situated inside the inner CJC.  The trajectories of charged particles are measured with a transverse momentum resolution of $\sigma
(p_T)/p_T \approx 0.2\% p_T/$GeV$\oplus 1.5\%$. The forward tracking detector (FTD) \cite{FTDJINST_FromKarin} measures the tracks of charged particles at polar angles \mbox{$6^{\circ} < \theta <  25^{\circ} $}. In the region of angular overlap, FTD and short CTD track segments are used to reconstruct combined tracks, extending the detector acceptance for well-reconstructed tracks. 
Both the CTD and the combined tracks are linked to hits in the vertex detectors: the central silicon tracker (CST) 
\cite{CST34,CST35}, the backward silicon tracker (BST) \cite{BST37} and the forward silicon tracker (FST) \cite{FST36}. 
These detectors provide precise spatial coordinate measurements and therefore significantly improve the primary vertex spatial resolution. 
The CST consists of two layers of double-sided silicon strip detectors surrounding the beam pipe covering an angular range of \mbox{$30^{\circ} < \theta  < 150^{\circ}$} for tracks passing through both layers. 
The BST consists of six double wheels of strip detectors measuring the transverse coordinates of charged particles. The FST design is similar to the BST and consists of five double wheels of single-sided strip detectors. 

The lead-scintillating fibre calorimeter (SpaCal) \cite{SPACAL}
covering the region \mbox{$153^{\circ} < \theta < 177.5^{\circ} $},
has electromagnetic and hadronic sections. The calorimeter is used to
measure the scattered positron and the backward hadronic energy flow.
The energy resolution for positrons in the electromagnetic section is
$\sigma (E)/E \approx 7.1\% / \sqrt{E/\text{GeV}} \oplus 1\%$, as
determined in test beam measurements \cite{test-beam-Spacal}.  The
SpaCal provides energy and time-of-flight information used for
triggering purposes. A backward proportional chamber (BPC) in front of
the SpaCal is used to improve the angular measurement of the scattered
lepton.  The liquid argon (LAr) calorimeter \cite{Lar} covers the range
\mbox{$4^{\circ} < \theta < 154^{\circ} $} and is used in this
analysis in the reconstruction of the hadronic final state.  It has an
energy resolution of $\sigma (E)/E \approx 50\% / \sqrt{E/\text{GeV}}
\oplus 2\%$ for hadronic showers, as obtained from test beam
measurements \cite{test-beam-Lar}.


\subsection{Event reconstruction}
The DIS kinematics is reconstructed based on the measurement of the scattered electron and the hadronic final state (HFS) particles. In the so-called $e\Sigma$-method
\cite{eSigma} the kinematic variables $Q^2$, $y$ and $x$ are given by:
\begin{equation} 
Q^2  = 4E_eE_e'\cos^2 \left( \frac{\theta_e}{2} \right), \quad
y = 2E_e \frac{\Sigma}{\left[\Sigma+E_e'(1-\cos\theta_e)\right]^2}\quad {\rm and}\quad x=\frac{Q^2}{s\cdot y},  
\end{equation}
where $s$ is the square of the centre-of-mass energy, $E_e'$ and $\theta_e$ the energy and polar angle of the scattered lepton, respectively,  $E_e$ being the energy of incoming lepton and  $\Sigma = \sum_i (E_i-p_{z,i})$ where the sum runs over all hadronic final state (HFS) particles $i$. This method provides an optimum in resolution of the kinematic variables and shows only little sensitivity to QED radiative effects.
The HFS particles are reconstructed using an energy flow algorithm \cite{hadroo2}. This
algorithm combines charged particle tracks and calorimetric energy clusters, taking
into account their respective resolution and geometric overlap, into hadronic
objects, while avoiding double counting of energy.

\subsection{Data selection}

DIS events were recorded using triggers based on electromagnetic energy deposits in the SpaCal calorimeter. The trigger efficiency is determined using independently triggered data. For DIS events the trigger inefficiency is negligible in the kinematic region of the analysis.

The scattered lepton, defined by the most energetic SpaCal cluster, is required to have an energy $E'_e$ larger than $12$ GeV. The kinematical phase space is defined by $5 < Q^2 <$ \mbox{$100$ GeV$^2$} and $0.05 < y < 0.6$, corresponding to the geometric acceptance of the SpaCal.  
The upper cut on $y$ reduces background from photoproduction. In addition, $x$ is required to be
in the range of $0.0001 < x < 0.01$. 

Additional selections are made to reduce QED radiation effects and to suppress background events. The $z$ coordinate of the event vertex is required to be within $35$ cm of the nominal interaction point. 
Events with high energy initial state photon radiation are rejected by requiring $35< \sum_i (E_i-p_{z,i})<75$ GeV.
Here, the sum extends over all HFS 
particles and 
the scattered electron. This cut further suppresses photoproduction background events to a level of about $0.5 \%$.


The tracks used in the analysis are measured in the CTD alone (central tracks) or result from combinations of CTD and FTD information (combined tracks). Central tracks are required to have transverse momenta in the laboratory frame $p_{T} > 150$ MeV. 
The momentum of a combined track is required to be larger than $0.5$ GeV to ensure that the track has enough momentum to cross the endwall of the CJC. Both central and combined tracks are required to originate from the primary event vertex and to be in the pseudorapidity range $-2< \eta < 2.5$ measured in the laboratory frame. Using only tracks assigned to the event vertex, the contributions from in-flight decays of $K_S^{0}$, $\Lambda $ and from photon conversions and from other secondary decays are reduced. Further track quality cuts \cite{myThesis} are applied to ensure a high purity of the track reconstruction.


 
\subsection{Definition of experimental observables}

The results of this analysis are presented in the hadronic centre-of-mass frame (HCM), to minimise the effect of the transverse boost from the virtual photon. The transformation to the HCM frame is reconstructed with the knowledge of the kinematic variables $Q^2$ and $y$ \cite{myThesis}. 
The transverse momentum and pseudorapidity of charged particles in the HCM frame are labelled as $p_T^*$ and $\eta^*$. 
Since in this frame the positive $z^*$ axis is defined by the direction of the virtual photon, 
HFS particles with $\eta^* > 0$ belong to the current hemisphere and particles with $\eta^* < 0$ 
originate from the target (proton remnant) hemisphere.

Charged particle densities as a function of transverse momentum and pseudorapidity are defined as $(1/N)({\rm d}n/{\rm d}p_T^*)$ and $(1/N)({\rm d}n/\rm{d}\eta^*)$, respectively. Here, ${\rm d}n$ is the total number of charged particles with transverse momentum (pseudorapidity) in the ${\rm d}p_T^*$ (${\rm d}\eta^*$) bin and $N$ denotes the number of selected DIS events. For distributions measured differentially in $x$ and $Q^2$, ${\rm d}n$ and $N$ are 
the numbers for the respective $(x,Q^2)$ bin. 

\begin{figure}[!h]
\begin{center}
\includegraphics[height=4.5cm,width=7.cm]{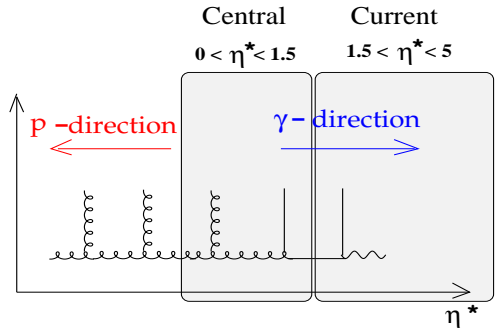}
\end{center}
  \caption{ The two pseudorapidity regions analysed in this paper. The region $0 < \eta^* < 1.5$ and  $1.5 < \eta^* <
5$, are denoted as ``central'' and ``current'' regions, respectively.}
  \label{fig:eta-feydiag} 
\end{figure}

Hadronisation is expected to be more relevant at small transverse momenta, while the hard parton radiation is expected to contribute more significantly at high $p_T^*$ ($p_T^* > 1$ GeV) \cite{Kuhlen-pt}. 
To distinguish hadronisations effects from parton evolution signatures, 
the charged particle density is measured as a function of  $\eta^*$ for \mbox{$0 < p_T^* < 1$ GeV} and for \mbox{$1 < p_T^* < 10$ GeV}. 
The $p_T^*$ dependence of the charged particle densities is studied in two different pseudorapidity intervals, $0 < \eta^* < 1.5$ and $1.5 <
\eta^* < 5$, referred to as the ``central region'' and ``current region'', respectively, as 
illustrated in \mbox{figure \ref{fig:eta-feydiag}}. Such division approximately defines the regions where the sensitivity to the hard scatter is largest (current region), and where the parton shower models can be tested (central region). The target region, $\eta^{*}<0$, is not accessible in this analysis.
%

\section{Data corrections}

The data are corrected to the number of stable charged particles including charged hyperons,
with proper lifetime $c\tau>10$ mm, in the phase space given in \mbox{table \ref{tab:sel-genlev}}.
Correction factors are calculated for each analysis bin from the
ratio of the number of generated stable charged particles to the
number of reconstructed tracks.
The bin widths are chosen such that a purity\footnote{The purity is
defined as the ratio of the number of charged particles generated
and reconstructed in a given bin to the total number of charged
particles in the phase space of the analysis which are reconstructed
in this bin.} of more than $75 \%$ is ensured in all bins.
The correction takes into account detector effects like limited
resolution and losses near the phase space boundaries, as well as a
small residual contamination from weak decays of neutral particles
(e.g. $K^0$ and $\Lambda$). 

\renewcommand{\arraystretch}{1.15}
\begin{table}[t]
\centering
\begin{tabular}{l|l} \hline \hline
\multicolumn{2}{c} {{ \bf DIS selection}}  \\ 
\hline
Four momentum transfers squared & $5 < Q^2 < 100$ GeV$^2$    \\
Inelasticity & $0.05 < y < 0.6$           \\
\hline
\multicolumn{2}{c} {{ \bf Charged particles}} \\
\hline 
Pseudorapidity in the laboratory frame  & $ -2< \eta < 2.5$ \\
Transverse momentum in the laboratory frame  & $p_{T} > 150$ MeV \\ 
Pseudorapidity in the HCM frame  & $ 0 < \eta^* < 5$ \\
Transverse momentum in the HCM frame  & $ 0 < p_T^*  < 10$ GeV \\ \hline \hline
\end{tabular}
\caption{Phase space for charged particles.}
\label{tab:sel-genlev}
\end{table}

In addition to migrations between bins inside the measurement phase space, there are migrations from outside of the analysis phase space and there is background from photoproduction. These contributions are subtracted prior to applying the correction factors according to the procedure outlined in \cite{forw-jet-h1-zeus3}.

The {\sc Djangoh} MC was used to correct the data. The differences to the correction factors obtained from {\sc Rapgap} are taken as systematic uncertainties. 
The correction factors strongly depend on $\eta^*$ and to a lesser extent on $p_T^*$. 
In the $1.5 < \eta^* < 5$ region they vary between $1$ and $1.8$ with
the largest values seen at high $p_T^*$ and large $\eta^*$. 
In the $0 < \eta^* < 1.5$ region, the correction factors rise up to $2.6$ at high $p_T^*$, due to the limited detector acceptance in this region. 
The two MC models predict very similar correction factors for most of the phase space region, but differences 
up to $5.5 \%$ are observed at small $\eta^*$ and high $p_T^*$.

\section{Systematic uncertainties}

The following sources of systematic uncertainties are considered for all measured quantities.

\begin{itemize}

\item The systematic uncertainty on the SpaCal energy scale is $1 \%$ \cite{ElCalib}, which results in a systematic uncertainty of typically $0.4 \%$ for the measured single differential distributions.

\item The SpaCal angular resolution of $1$ mrad leads to a systematic uncertainty of about $0.1 \%$ for the measured distributions.

\item The hadronic energy scale uncertainty is known to a precision of $2\%$ \cite{HFSCalib}. The scale uncertainty enters into the uncertainty of the phase space calculations, which depend on $E-P_z$ of the HFS, and also affects the boost to the HCM frame. The systematic effect on the present measurements is about $0.3\%$.

\item The systematic uncertainty arising from the model dependence of
  the data correction is taken as the difference of the correction
  factors calculated using {\sc Rapgap} and {\sc Djangoh} MC. The
  resulting uncertainty on the measurements varies between $0.2 \%$
  and $5.5 \%$.

\item The systematic uncertainty associated with the track reconstruction (e.g. track  reconstruction efficiency, vertex reconstruction efficiency, weak decays and nuclear interaction uncertainties) is estimated to be:

\begin{itemize}
\item $1 \%$ for central tracks, determined from the analysis of curling tracks and from the analysis of secondary vertices of charged particles located in the material between the two CJCs and originating from 
interactions with the detector material. 
The nuclear interaction cross sections of pions and kaons is found to be smaller in the simulation than in data. After correcting for these deficits, the agreement in the track efficiency between data and MC is found to be better than $1\%$. 
\item $10 \%$ for combined tracks \cite{myThesis, FTDJINST_FromKarin}. This was checked using all selected central tracks, as well as by using pions from $K_S^0$ decays, as a function of transverse momentum and pseudorapidity. Consistent results are obtained from both samples showing agreement of data and MC within $10 \%$. 
\end{itemize}

The systematic uncertainty associated with the track reconstruction is applied as an independent uncorrelated uncertainty on every data point. The resulting uncertainty on the measurements varies between $1 \%$ and $5.4 \%$ and is $1.6 \%$ on average. An additional systematic uncertainty of $0.2\%$ is assigned due to the different $K_S^0$ contamination seen in data and MC for both central and combined tracks. The corresponding effect arising from 
$\Lambda$ contamination is expected to be negligible. 

\item The systematic uncertainty on the remaining photoproduction background is estimated to be $30\%$. This results in an uncertainty on the measured densities up to $0.9 \%$ at small  $x$ and $Q^2$. At large $x$ and $Q^2$ the contribution from photoproduction is small and its uncertainty is negligible.

\end{itemize}

The systematic uncertainties shown in the figures and tables 
are calculated by adding all contributions in quadrature. 
The total systematic uncertainty for the single differential measurements is 
below  $2.5\%$ for most analysis bins.

\section{Results}

The measurements of the charged particle densities as a function of pseudorapidity and transverse momentum in the phase space summarised in \mbox{table \ref{tab:sel-genlev}} 
are listed in tables \ref{tab:pt-curr} to \ref{tab:pt-curr7} and shown in figures \ref{fig:eta-PDF-RAPGAP} to \ref{fig:pt-bins-curr}.

\subsection{Charged particle densities as a function of pseudorapidity}

\begin{figure}[!p]
 \centering
    \unitlength1cm
         \begin{picture}(18.,8.5)(0.0,0.0) 
\put(-0.0,0.){\includegraphics*[height=8.8cm,width=8.45cm]{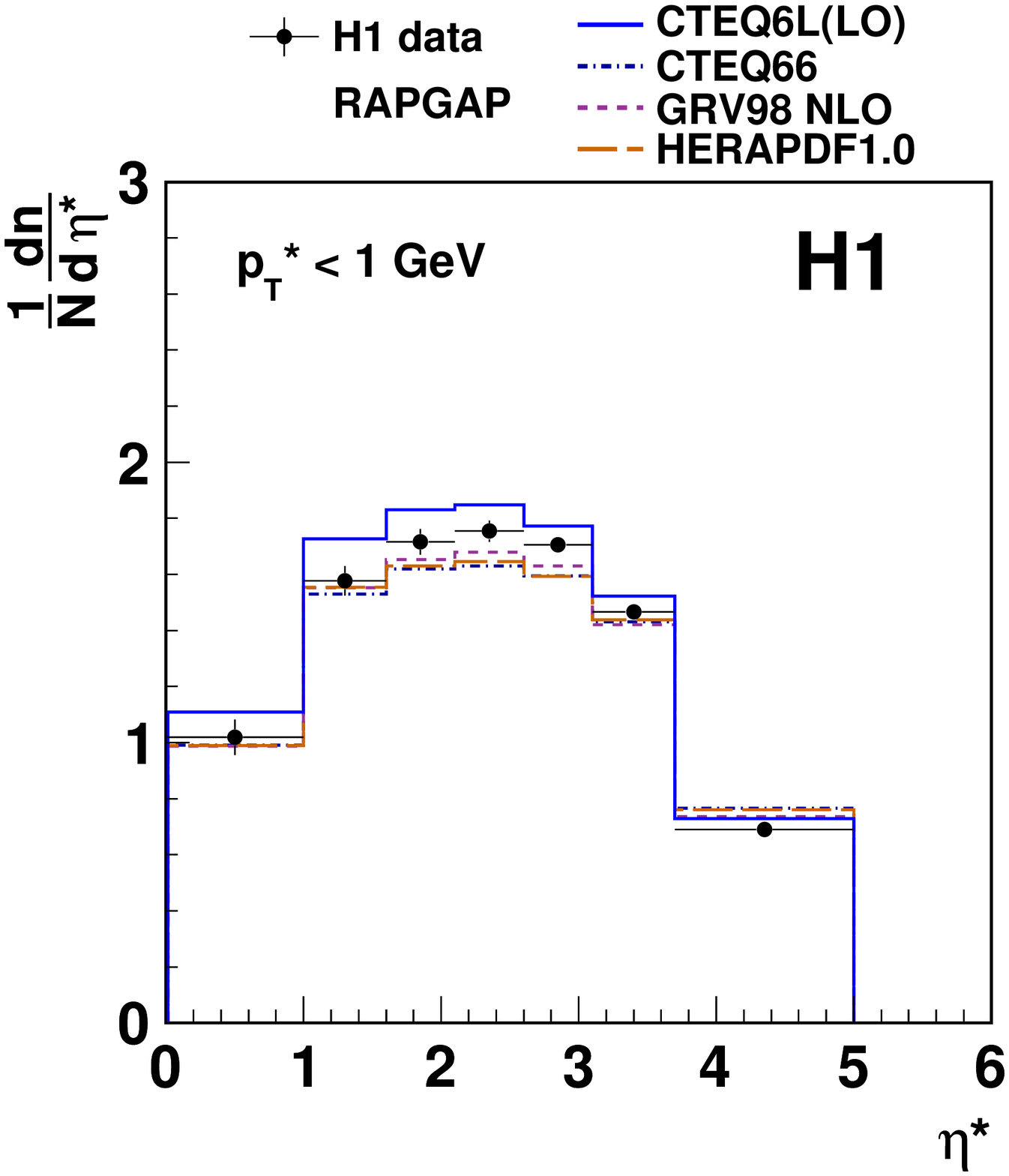}}
\put(8.0,0.){\includegraphics*[height=8.8cm,width=8.45cm]{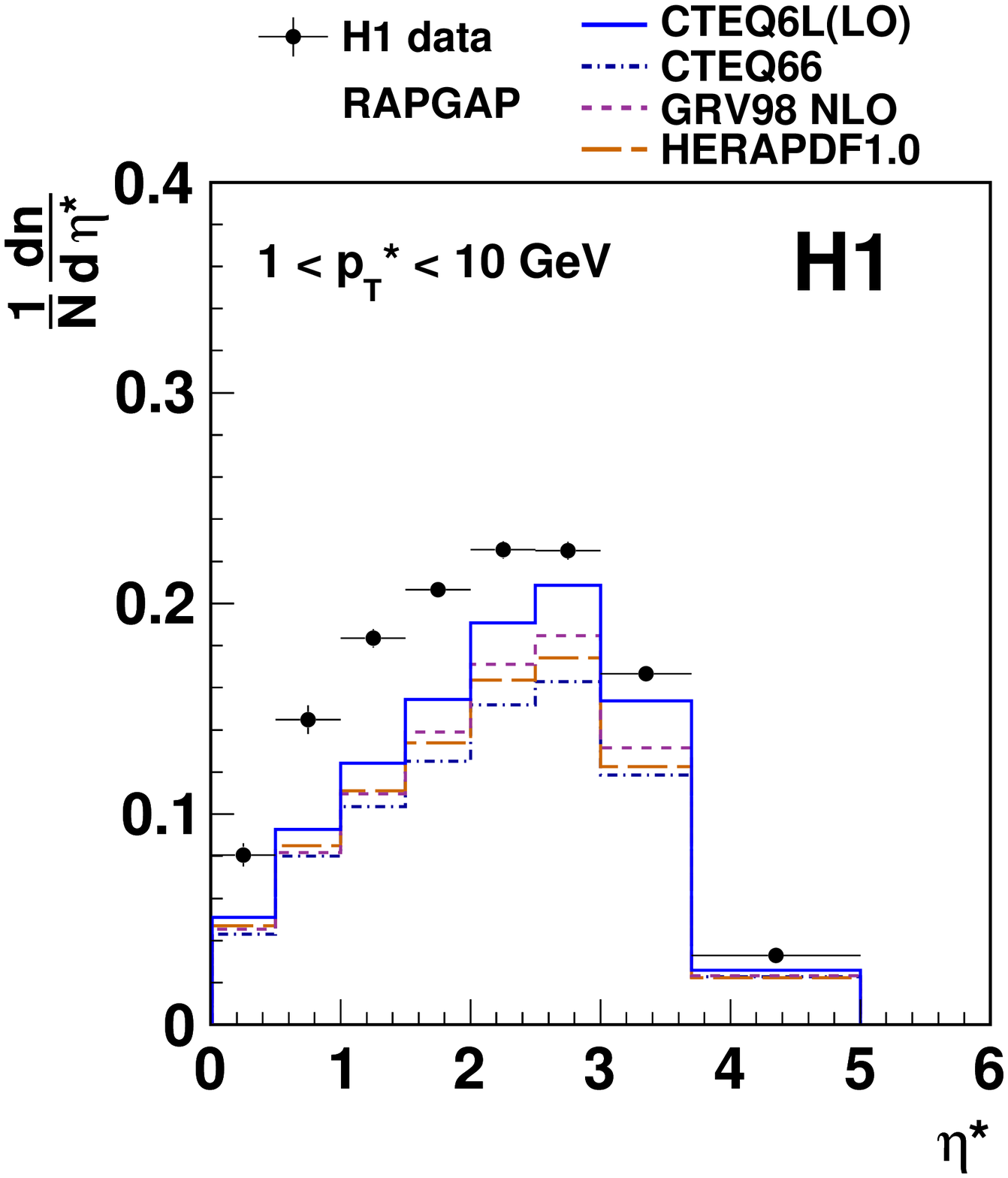}}
\put(0.6,8.2){{ \bf \Large{(a)} }}
\put(8.7,8.2){{ \bf \Large{(b)} }}
\end{picture}
\caption{Charged particle density as a function of $\eta^*$ for (a) \mbox{$p_T^*< 1$ GeV} and for (b)
\mbox{$1< p_T^*< 10$ GeV} compared to {\sc Rapgap} predictions with different proton PDFs. The predictions are obtained using the ALEPH tune.}
\label{fig:eta-PDF-RAPGAP} 

\end{figure}

The charged particle densities as a function of $\eta^*$ were measured separately for \mbox{$p_T^* < 1$ GeV} and for \mbox{$1 < p_T^* < 10$ GeV}, as shown in \mbox{figure
  \ref{fig:eta-PDF-RAPGAP}}. In the soft $p_T^*$ region, the pseudorapidity distribution is almost flat in the $1.5 \lesssim \eta^* \lesssim 3$ range with about $1.7$ charged particles per unit of pseudorapidity. The distribution falls at small $\eta^{*}$ due to the cut on pseudorapidity in the laboratory frame. In the hard $p_T^*$ region the distribution becomes more peaked near $\eta^{*}=2.5$, with a maximum of $0.23$ charged particles per unit of pseudorapidity. For $1 < p_T^* < 10$ GeV the density increases rather strongly up to 
$\eta^{*} \approx 2.5$, a behaviour expected from the strong ordering of transverse momentum towards the hard scattering vertex.

\mbox{Figure \ref{fig:eta-PDF-RAPGAP}} also shows the predictions of the DGLAP-like model {\sc Rapgap} based on different PDF sets. In the soft $p_T^*$ region all NLO PDFs (HERAPDF1.0 \cite{HERAPDF10}, CTEQ6.6 \cite{CTEQ66M}, GRV98NLO \cite{GRV98}) show similar results and predict less particles compared to calculations using the default LO PDF set CTEQ6L(LO). All predictions are close to the data. At large $p_T^*$, differences between the NLO PDF sets are observed, with CTEQ6L(LO) being closest to the data, 
although the differences to the data are still larger than the differences between the various PDF predictions. Similar PDF 
uncertainties are observed when 
using the CDM model as implemented in {\sc Djangoh}.

\begin{figure}[!b]
\vspace{0.6cm}
 \centering
    \unitlength1cm
         \begin{picture}(18.,8.5)(0.0,0.0) 
\put(-0.0,0.){\includegraphics*[height=8.8cm,width=8.45cm]{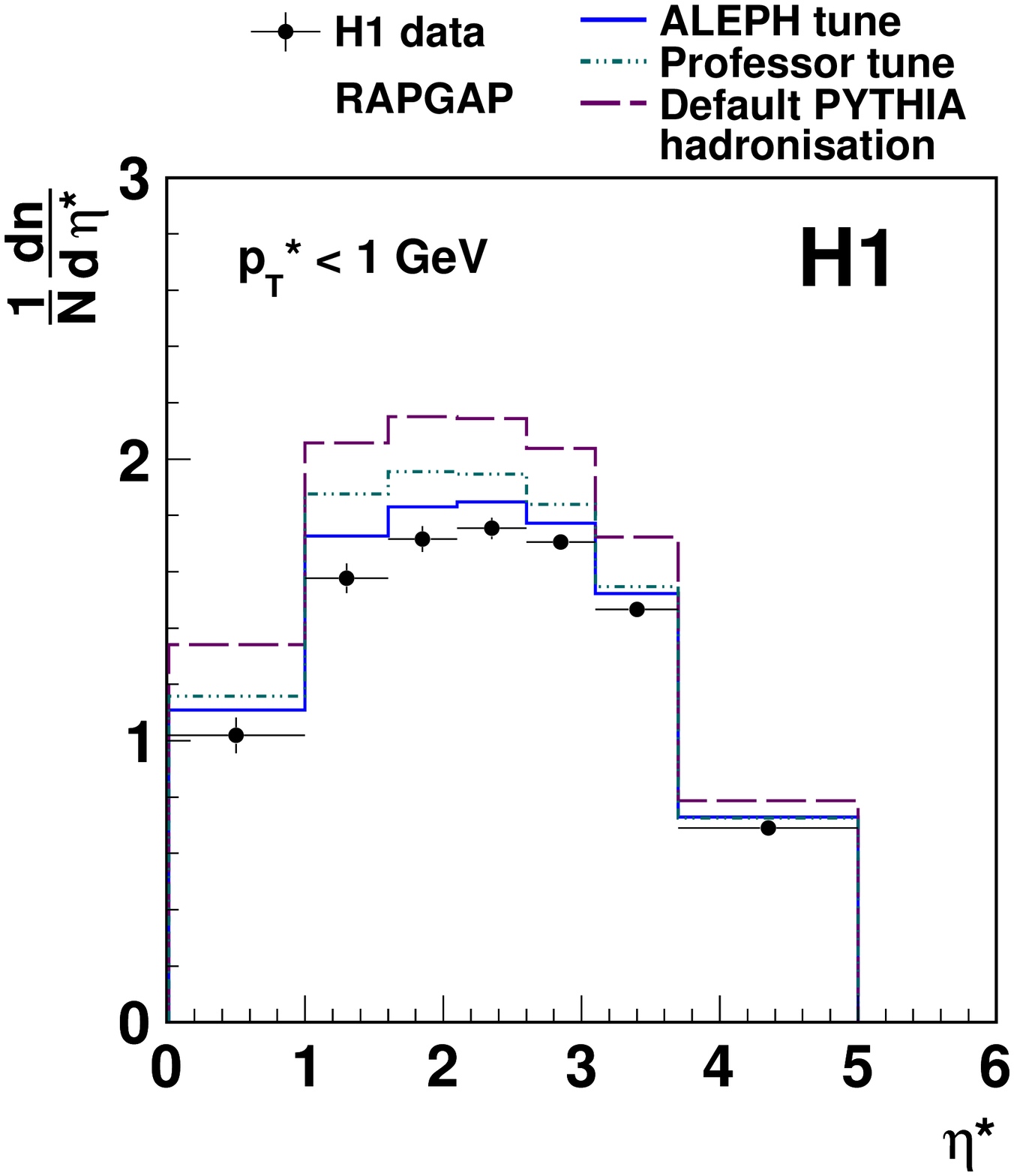}}
\put(8.0,0.){\includegraphics*[height=8.8cm,width=8.45cm]{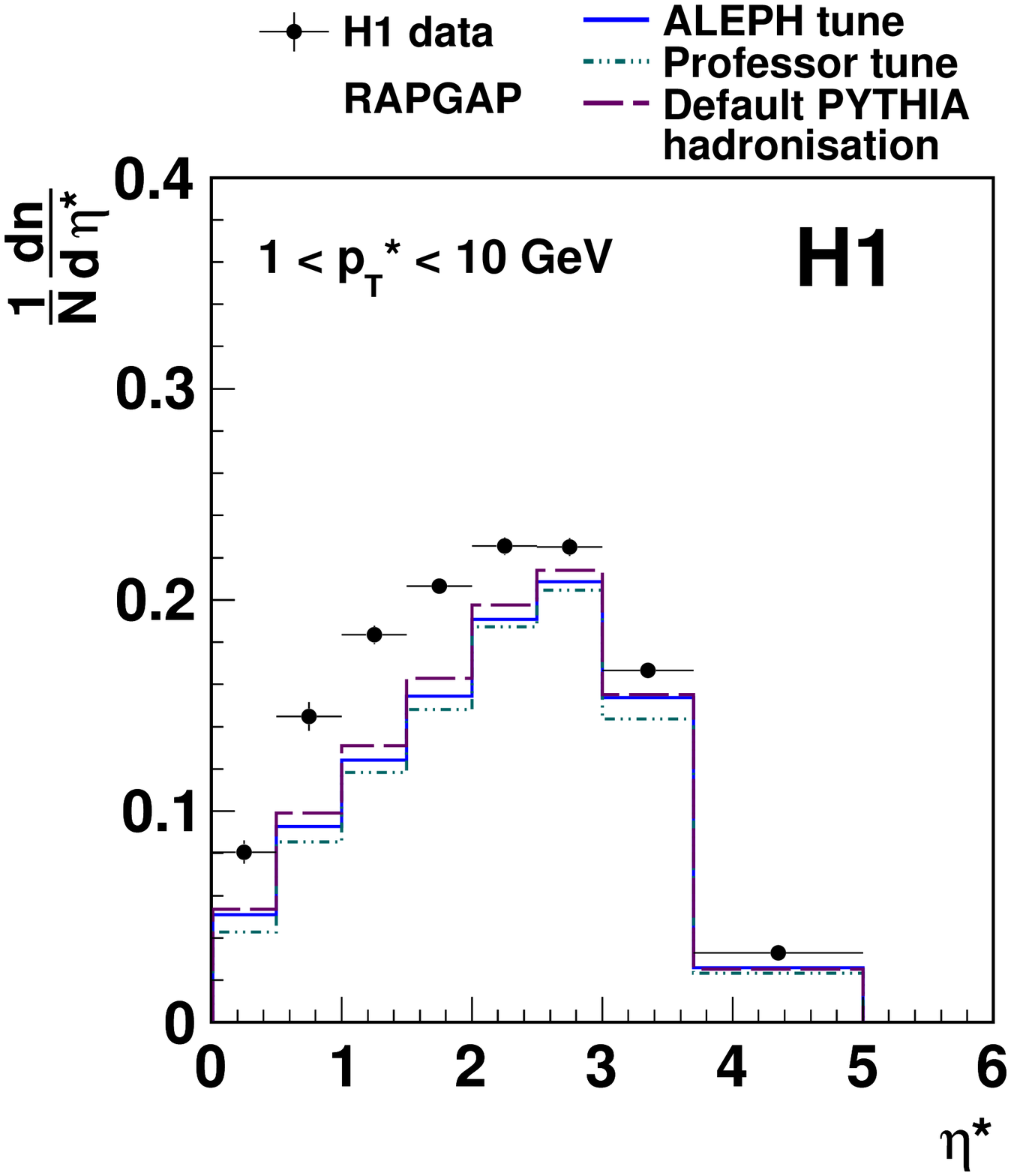}}
\put(0.6,8.2){{ \bf \Large{(a)} }}
\put(8.7,8.2){{ \bf \Large{(b)} }}
\end{picture}
\caption{Charged particle density as a function of $\eta^*$ for (a) $p_T^*< 1$ GeV for (b) \mbox{$1< p_T^*< 10$ GeV} compared to {\sc Rapgap} predictions for three different sets of fragmentation parameters. The predictions are obtained using CTEQ6L(LO) PDF.}
\label{fig:eta-aleph-prof-def} 
\end{figure}

To check the sensitivity to hadronisation effects, the {\sc Rapgap} predictions obtained with three sets of fragmentation parameters are compared to the data in \mbox{figure \ref{fig:eta-aleph-prof-def}}: parameters tuned by ALEPH \cite{ALEPH}, by the Professor tuning tool \cite{profftune} and default {\sc Pythia}6.424 fragmentation parameters. Significant differences between these three samples are seen in the soft $p_T^*$ region, where the data are best described by the ALEPH tune. At large transverse momenta they give similar results but none of them describes the data.

\begin{figure}[!b]
 \centering
    \unitlength1cm
         \begin{picture}(18.,9.)(0.0,0.0) 
\put(-0.0,0.){\includegraphics*[height=8.8cm,width=8.45cm]{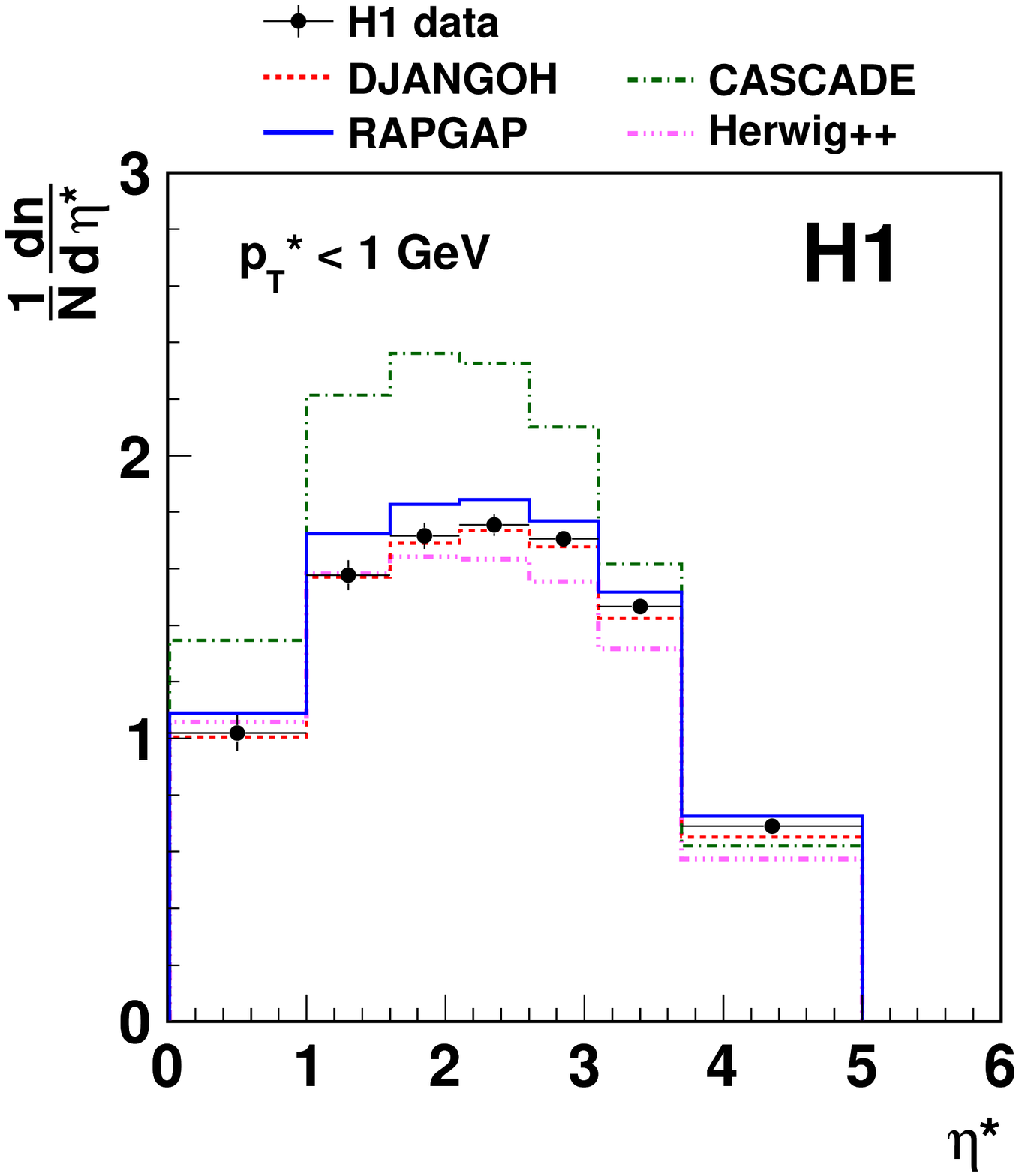}}
\put(8.0,0.){\includegraphics*[height=8.8cm,width=8.45cm]{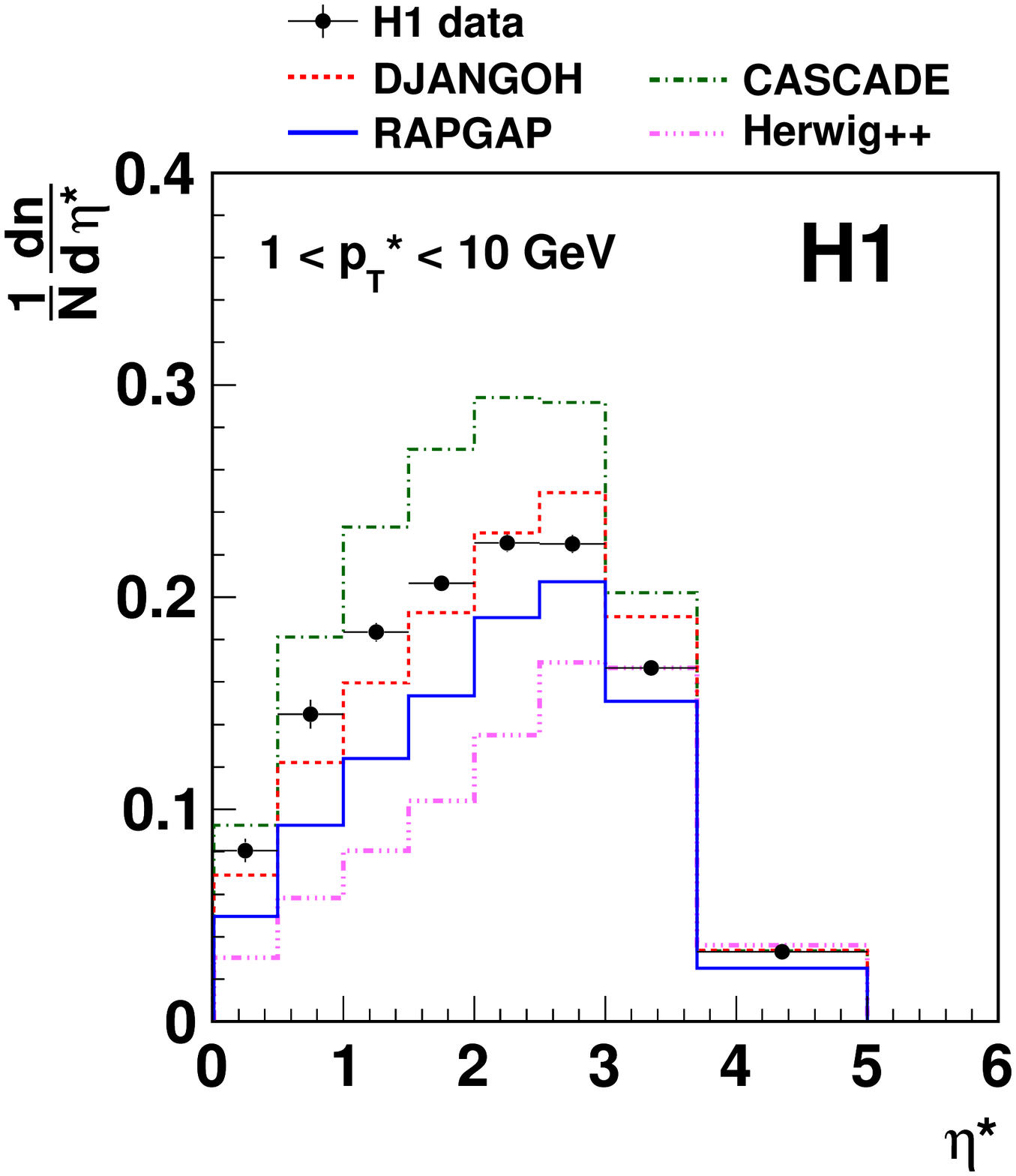}}
\put(0.6,8.2){{ \bf \Large{(a)} }}
\put(8.7,8.2){{ \bf \Large{(b)} }}
\end{picture}
\caption{Charged particle density as a function of $\eta^*$ for (a) $p_T^*< 1$ GeV for (b) \mbox{$1< p_T^*< 10$ GeV} compared to {\sc Djangoh}, {\sc Rapgap}, Herwig++ and {\sc Cascade} Monte Carlo predictions.} 
\label{fig:eta-PS} 
\end{figure}

Predictions from models with different approaches for QCD radiation (see section 2) are shown in \mbox{figure \ref{fig:eta-PS}}. The data are compared to the CDM model {\sc Djangoh}, the DGLAP-based MC {\sc Rapgap} and Herwig++  and the CCFM model {\sc Cascade}. In the soft $p_T^*$ region, {\sc Djangoh} and {\sc Rapgap} describe the data within the PDF uncertainties (figure \ref{fig:eta-PDF-RAPGAP}). Herwig++, which uses the cluster fragmentation model, provides a reasonable description of the data in the central region. The effect of not using the POWHEG option in Herwig++ also has been investigated. Only small differences were observed which are not considered further in this paper. {\sc Cascade} predicts too high multiplicities in most of the measured $\eta^*$ range.
In the region of $1< p_T^{*} <10$~GeV the best description of the data is achieved by {\sc Djangoh}. {\sc Rapgap} strongly undershoots the data in the central region.  Herwig++ 
predicts a spectrum which is even below the prediction of {\sc Rapgap}. {\sc Cascade} is significantly above the data in a wide range of $\eta^*$.

\begin{figure}[p]
\vspace{2cm}
    \unitlength1cmth
\hspace{-25mm}
         \begin{picture}(17.,18.)(0.0,0.0) 
     \includegraphics*[height=18.cm,width=17.cm]{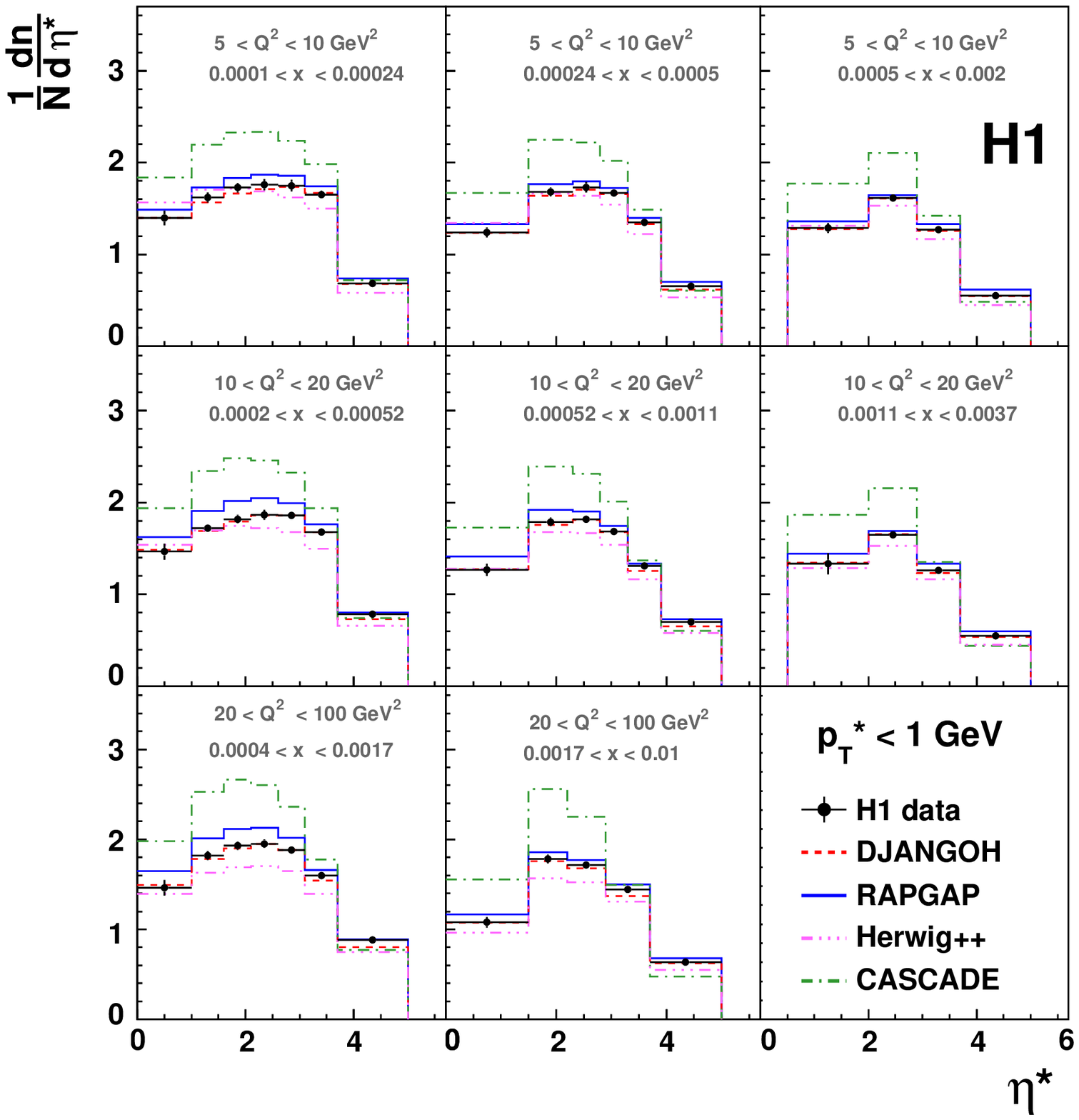}
\end{picture}
\vspace{-30mm}
\caption{Charged particle density as a function of $\eta^*$ for $p_T^*< 1$ GeV for eight 
intervals of $Q^{2}$ and $x$ compared to {\sc Djangoh}, {\sc Rapgap}, Herwig++ and {\sc Cascade} Monte Carlo predictions.}
\label{fig:eta-soft} 
\end{figure}

The charged particle densities as a function of $\eta^*$ are shown in \mbox{figure \ref{fig:eta-soft}} for \mbox{$p_T^* < 1$ GeV} in eight different intervals of $x$ and $Q^{2}$ . The data are compared to predictions of the {\sc Djangoh}, {\sc Rapgap}, Herwig++ and {\sc Cascade} generators. {\sc Djangoh} provides a good description of the data over the
full kinematic range. 
In general the description of the data by {\sc Rapgap} is somewhat worse, with overshooting the data by about 
$10\%$ at low $x$. 
Herwig++ predicts smaller charged particle densities than observed in data in most of the phase space with differences of the order of $10\%$ at the highest $Q^2$. 
{\sc Cascade} is significantly above the data for $\eta^{*} < 3$ in all $(x,Q^2)$ bins.

\begin{figure}[p]
\vspace{2cm}
    \unitlength1cm
\hspace{-25mm}
         \begin{picture}(17.,18.)(0.0,0.0) 
     \includegraphics*[height=18.cm,width=17.cm]{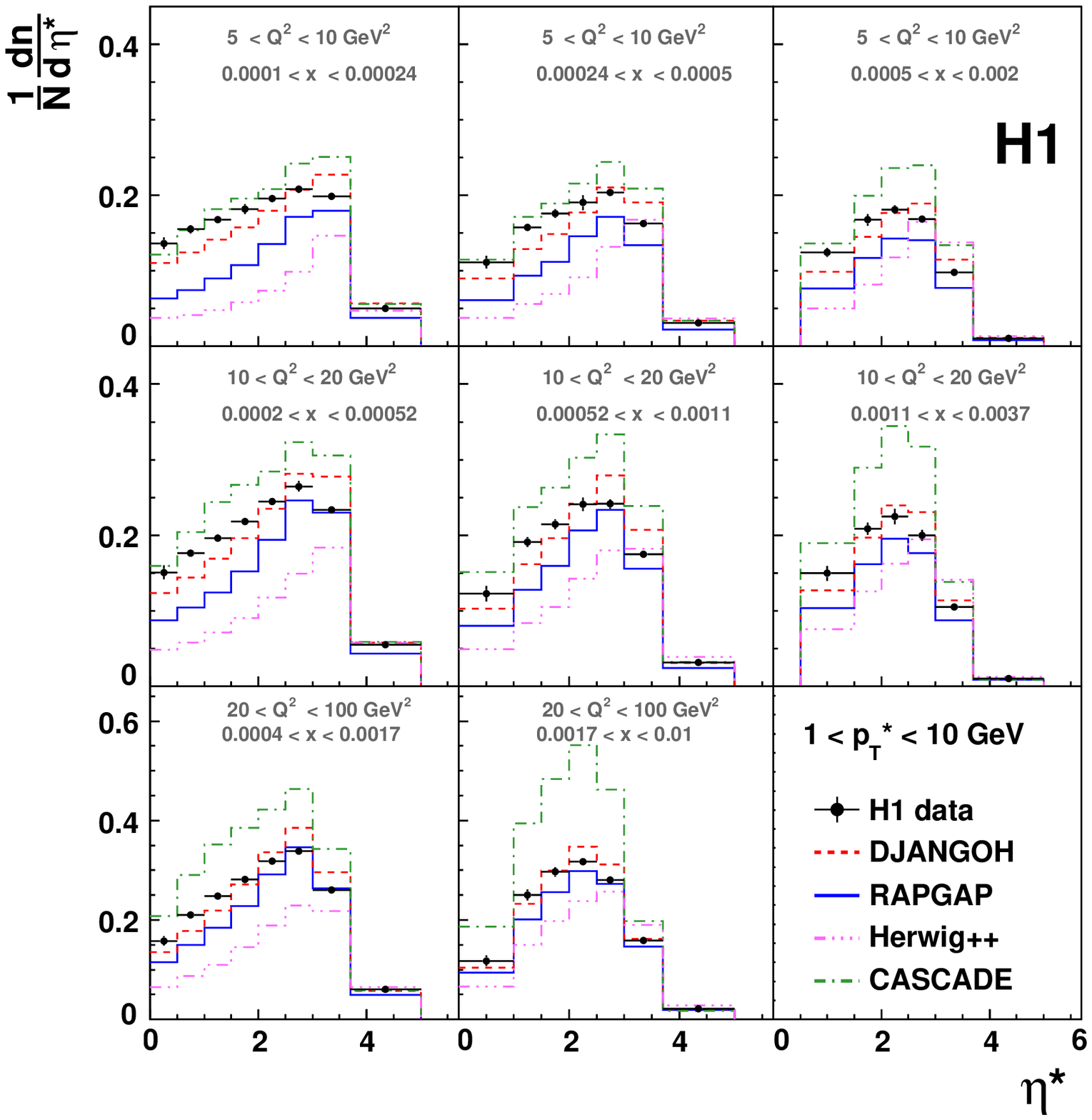}
\end{picture}
\vspace{-30mm}
\caption{Charged particle density as a function of $\eta^*$ for $1< p_T^*< 10$ GeV for eight 
intervals of $Q^{2}$ and $x$ compared to {\sc Djangoh}, {\sc Rapgap}, Herwig++ and {\sc Cascade} Monte Carlo predictions.}
\label{fig:eta-hard} 
\end{figure}

In figure  \ref{fig:eta-hard} the charged particle densities as a function of $\eta^*$ are shown in $(x,Q^{2})$ intervals for  $1< p_T^* < 10$ GeV . The shape of the distributions changes with $x$ and $Q^{2}$ more strongly than what is observed for \mbox{$p_T^* < 1$ GeV} (\mbox{figure \ref{fig:eta-soft}}). At small values of $x$ and $Q^{2}$ the measured distribution is less dependent on 
$\eta^{*}$ compared to the region at high $x$ and $Q^{2}$. None of the models describes all aspects of the data. In general {\sc Djangoh} is closest to the data. However it fails to 
describe the data at low and medium $x$ in the central pseudorapidity region, with downwards deviations of the order of $20\%$. The  {\sc Rapgap} prediction is below the data, with the strongest deviation observed at small $x$ and small $\eta^*$. Herwig++ significantly undershoots the data. The prediction of {\sc Cascade} agrees reasonably well with the measurement at low $x$ and $Q^2$, but overshoots the data significantly as $x$ or $Q^2$ increases.

\subsection{Charged particle densities as a function of transverse momentum}
\label{chap-result}

\begin{figure}[!h]
    \unitlength1cm
         \begin{picture}(18.,10.5)(0.0,0.0) 
\put(-0.3,0.){\includegraphics*[height=10.4cm,width=9.4cm]{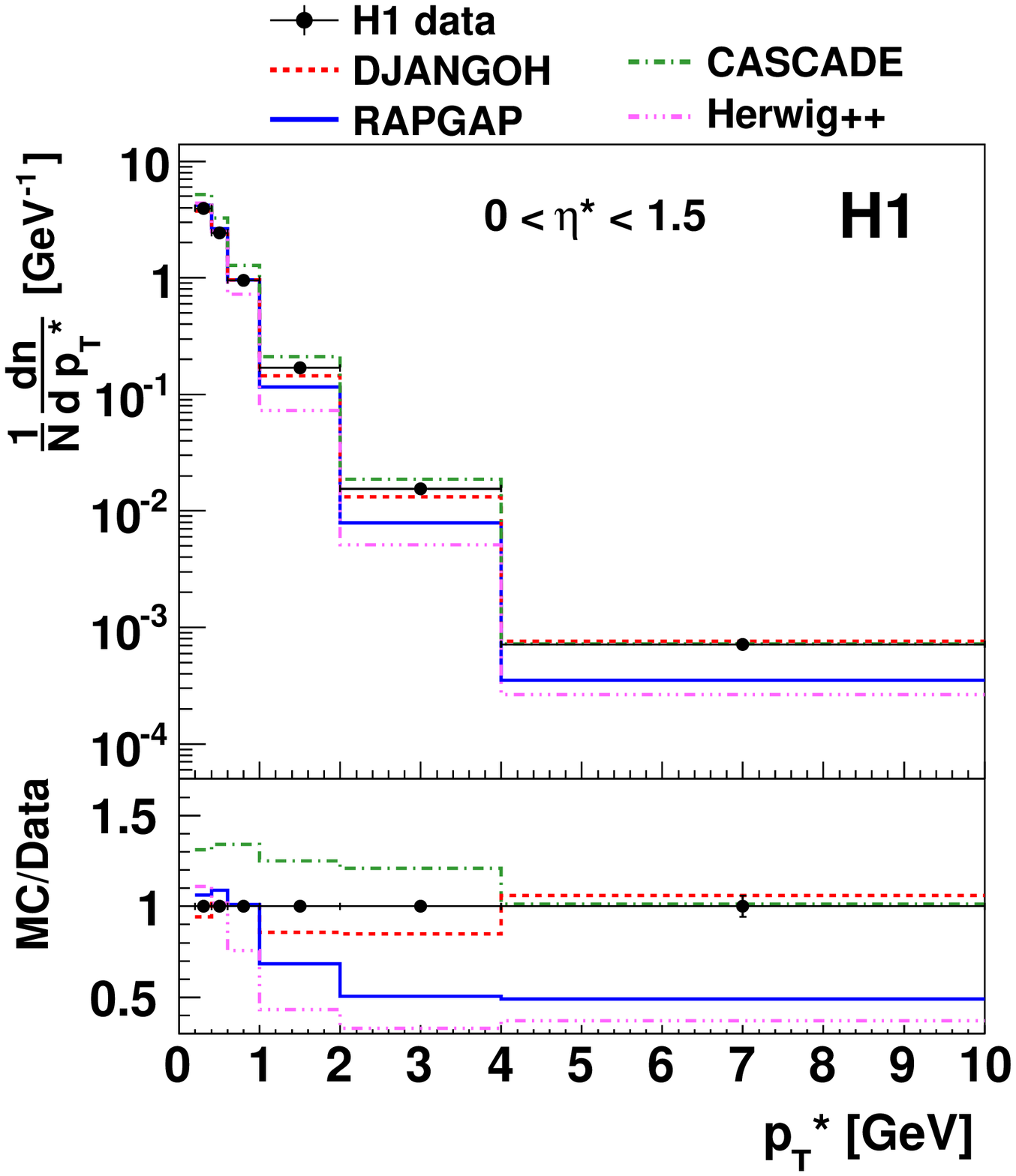}}
\put(7.8,0.){\includegraphics*[height=10.4cm,width=9.4cm]{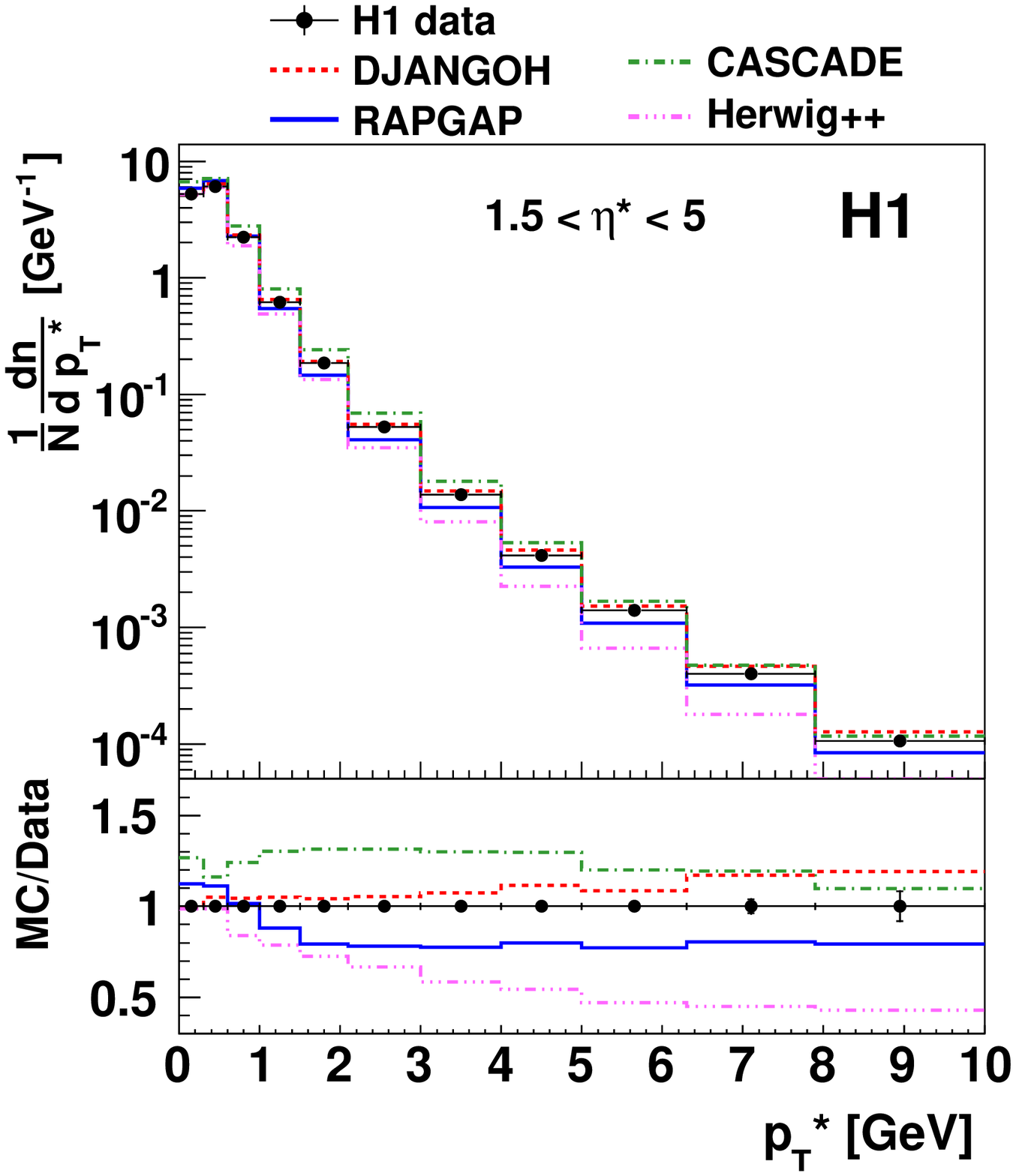}}
\put(0.6,9.2){{ \bf \Large{(a)} }}
\put(8.7,9.2){{ \bf \Large{(b)} }}
\end{picture}
\vspace{-1.0cm}
\caption{Charged particle density as a function of $p_T^*$ in the ranges (a) $0 <\eta^*< 1.5$ and (b) $1.5 <\eta^*< 5$ compared to {\sc Djangoh}, {\sc Rapgap}, Herwig++ and {\sc Cascade} Monte Carlo predictions. The ratios of MC predictions to the measurements are shown on the bottom of the figure.}
\label{fig:pt-mcdata-cen-RDC} 

\end{figure}

In \mbox{figure \ref{fig:pt-mcdata-cen-RDC}} the charged particle densities as a function of 
$p_T^*$ are shown for two pseudorapidity intervals, $0 < \eta^* < 1.5$ (central) and $1.5 < \eta^* < 5$ (current). The shapes of the measured $p_T^*$ distributions in the two pseudorapidity ranges are similar. The spectrum falls by more than four orders of magnitude from $p_T^* < 1$ GeV to $p_T^* \approx 8$ GeV. The measurements are compared to the predictions of the {\sc Djangoh}, {\sc Rapgap}, Herwig++ and {\sc Cascade} generators. {\sc Djangoh} provides in general a good description of the data, while only  
at high $p_{T}^*$ in the current region deviations from the measurement are observed. The other models
fail to describe the data, with the strongest deviations being observed in the central region.
The ratio of RAPGAP to data shows a sharp drop at $p_T^*\approx 1$ GeV.
The $p_T^*$ spectra predicted by Herwig++ are even softer than those predicted by {\sc Rapgap}.
{\sc Cascade} in general produces higher particle densities than measured.

\begin{figure}[p]
\vspace{2cm}
    \unitlength1cm
\hspace{-25mm}
         \begin{picture}(17.,18.)(0.0,0.0) 
     \includegraphics*[height=18.cm,width=17.cm]{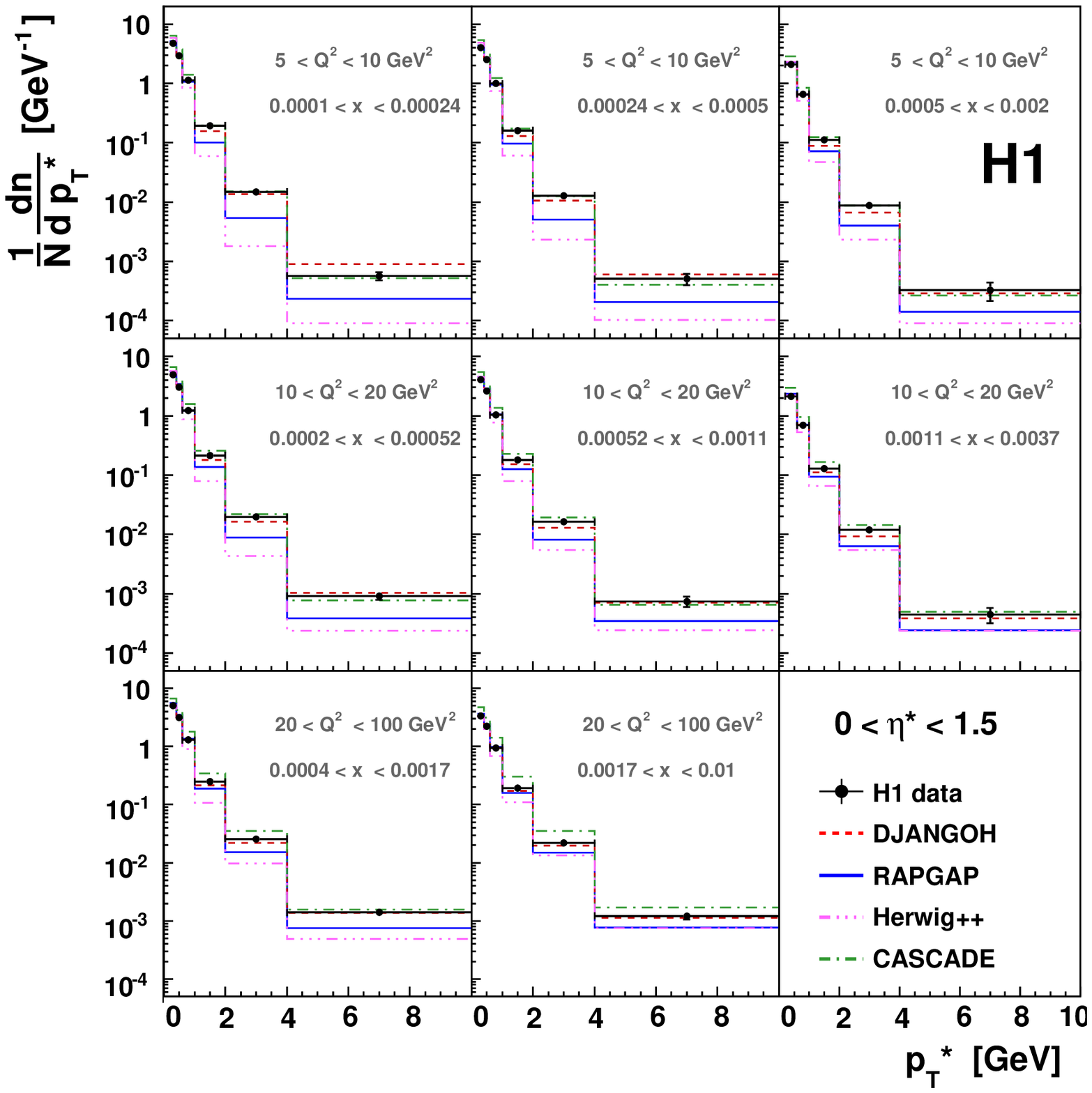}
\end{picture}
\vspace{-30mm}
\caption{Charged particle density as a function of $p_T^*$ in the range
 $0 <\eta^*< 1.5$ for eight intervals of $Q^{2}$ and $x$ compared to {\sc Djangoh}, {\sc Rapgap}, Herwig++ and {\sc Cascade} Monte Carlo predictions.}
\label{fig:pt-bins-cen} 
\end{figure}

\begin{figure}[p]
\vspace{2cm}
    \unitlength1cm
\hspace{-25mm}
         \begin{picture}(17.,18.)(0.0,0.0) 
     \includegraphics*[height=18.cm,width=17.cm]{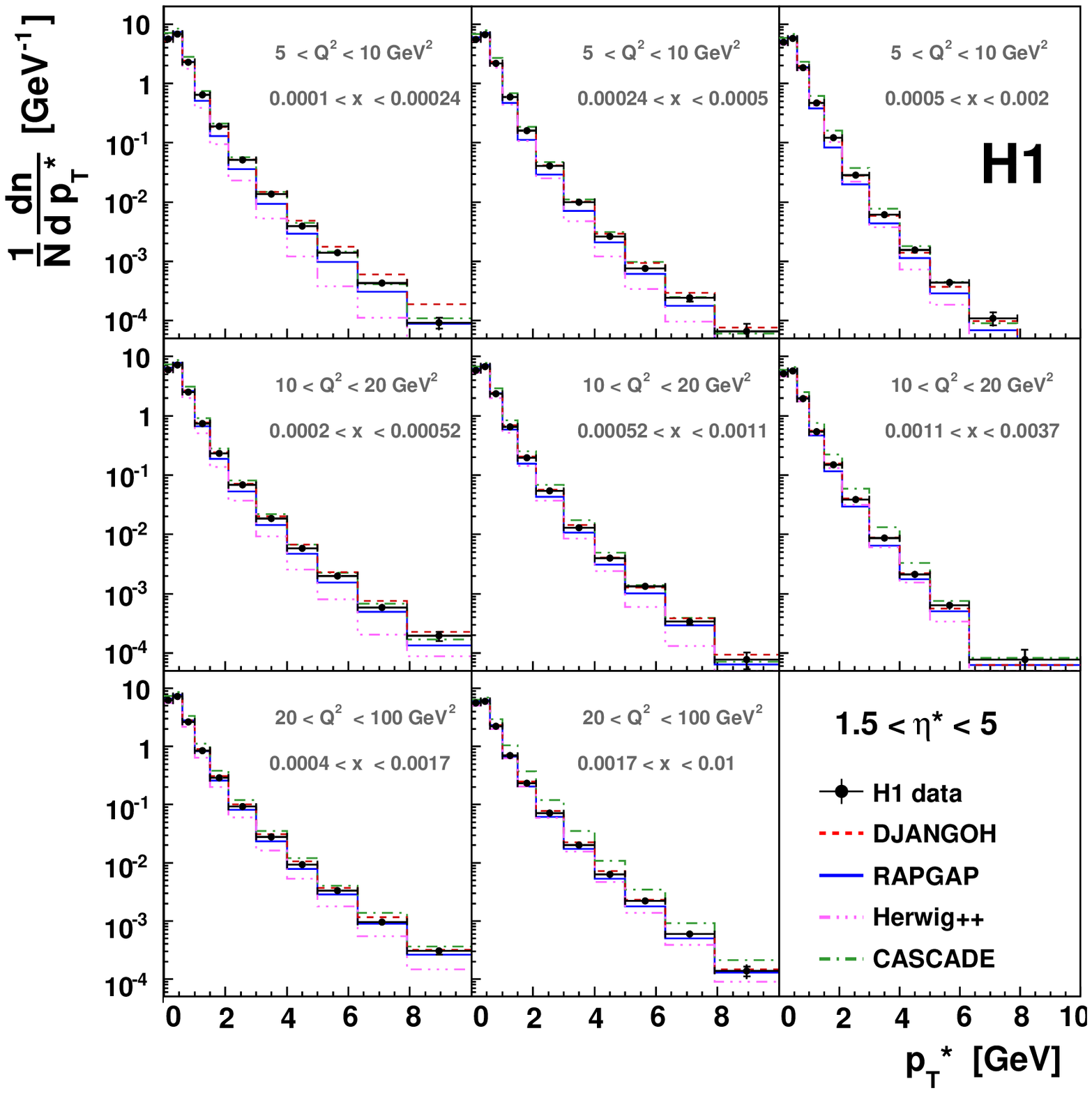}
\end{picture}
\vspace{-30mm}
\caption{Charged particle density as a function of $p_T^*$ in the range 
$1.5 <\eta^*< 5$ for eight intervals of $Q^{2}$ and $x$ compared to {\sc Djangoh}, {\sc Rapgap}, 
Herwig++ and {\sc Cascade} Monte Carlo predictions.}
\label{fig:pt-bins-curr} 
\end{figure}

In \mbox{figures \ref{fig:pt-bins-cen} and \ref{fig:pt-bins-curr}} the
charged particle densities as a function of $p_T^*$ are shown for eight $(x,Q^{2})$ intervals for the central and the current region, respectively. In the central region the measurement shows a
dependence on $x$, such that the number of soft particles is decreasing with increasing 
$x$ for fixed $Q^{2}$. In the current region this effect
is less pronounced. The {\sc Djangoh} model provides in general a good description of the data over the full kinematic range in both pseudorapidity regions, degrading at high $p_{T}^*$ in the lowest $(x,Q^{2})$ bin.
Significant deviations of the {\sc Rapgap} predictions from data
are observed in the central region at low $x $ and $Q^{2}$. The description becomes somewhat better at larger values of $x$ and
$Q^{2}$. The same trend is observed for the current region, but the
overall data description is better. Herwig++ fails to describe 
the measurements at high $p_{T}^*$ in the whole phase space. At 
lowest $x$ and $Q^{2}$ the spectrum is much softer than the one obtained with {\sc Rapgap}, while at high $x$ and $Q^{2}$ 
both predictions are similar.
{\sc Cascade} describes the data in the
lowest $x$ and $Q^{2}$ bin at high $p_{T}^*$ only.

\section{Conclusion}

This paper presents a study of charged particle production in $ep$ collisions 
at low $Q^2$ measured with the H1 detector. The kinematic range of the analysis covers low photon virtualities, \mbox{$5 < Q^2 < 100$ GeV$^2$}, and small values of $x$, $10^{-4} < x < 10^{-2}$.
The analysis is performed in the hadronic centre-of-mass system. The charged particle densities as a function of pseudorapidity ($\eta^*$) and transverse momentum ($p_T^*$) are measured differentially in $x$ and $Q^2$. The charged particle densities as a function of pseudorapidity show different shapes, depending on the $p_T^*$ range. For $0<p_T^* < 1$ GeV the density of particles is approximately constant for $1<\eta^*<3.5$, while for \mbox{$1< p_T^* < 10$ GeV} the density increases with increasing $\eta^{*}$ up to $\eta^* \approx 2.5$, a behaviour
expected from the strong ordering of transverse momentum towards the hard scattering vertex.
The charged particle densities as a function of transverse momentum
show an $x$ dependence at small $\eta^*$ ($0 < \eta^* < 1.5$), such that the number of soft 
particles is decreasing with increasing $x$, while in the $1.5 < \eta^* < 5$ range 
this effect is less visible.

In order to relate the charged hadron spectra to the parton dynamics at small $x$,
the data are compared to QCD models with different evolution
approaches for simulating the parton cascade and with different hadronisation 
schemes. The data allow the validity of different models to be tested.
At small $p_T^*$, the data 
are reasonably well described by {\sc Djangoh} (based on the Colour Dipole Model), as well as by 
{\sc Rapgap} (based on the DGLAP shower evolution). At high $p_T^*$ and at low $\eta^*$,
{\sc Rapgap} severely undershoots the data. The differences are most pronounced
at lowest $x$ and $Q^2$, and decrease with increasing $x$ and $Q^2$ values.
Herwig++ which is also based on DGLAP but uses a cluster fragmentation model is significantly
below the data over the full phase space. {\sc Cascade} (based on CCFM) gives a reasonable description only at the lowest $x$ and $Q^2$, but overall predicts higher charged particle densities than observed in data. The Colour Dipole Model implemented in {\sc Djangoh} is the best among the 
considered models and provides a reasonable description of the data.


\section{Acknowledgement}
We are grateful to the HERA machine group whose outstanding efforts have made this experiment
possible. We thank the engineers and technicians for their work in constructing and
maintaining the H1 detector, our funding agencies for financial support, the DESY technical
staff for continual assistance and the DESY directorate for support and for the hospitality which
they extend to the non-DESY members of the collaboration.
We would like to give credit to all partners contributing to the WLCG
computing infrastructure for their support for the H1 Collaboration.


\clearpage





\clearpage



\renewcommand{\arraystretch}{1.15} 
\begin{table}[h]
  \begin{center}
   \tabcolsep 5.8pt
    \normalsize
    \begin{tabular}{|c|ccc|}
      \hline
     \multicolumn{4}{|c|} {\bf \boldmath 
     $0<p_T^* <$ 1 GeV }  \\
     \hline
     \hline
   \multicolumn{1}{|c|}  {$\eta^*$ range} & $1\!/\!N \cdot {\rm d}n/{\rm d}\eta^*\,$ & stat. & sys.  \\  
    \multicolumn{1}{|c|}  {} & \multicolumn{1}{c}{} &\multicolumn{1}{c}{($\%$)} & \multicolumn{1}{c|}{($\%$)} \\ 
      \hline
                                                   $0.0 - 1.0$  & $1.019$  & $0.06$   &  $6.2$       \\ 
                                                   $1.0 - 1.6 $ & $1.577$  & $0.03$   &  $3.4$       \\ 
                                                   $1.6 - 2.1$  & $1.717$  & $0.03 $  &  $2.7$      \\ 
                                                   $2.1 - 2.6$  & $1.754$  & $0.03$   &  $2.1$      \\ 
                                                   $2.6 - 3.1$  & $1.706$  & $0.03$   &  $1.4$       \\   
                                                   $3.1 - 3.7$  & $1.467$  & $0.04$   &  $1.4$       \\ 
                                                   $3.7 - 5.0 $ & $0.691$  & $0.08 $  &  $1.7$       \\  
     \hline
    \end{tabular}
    \caption{Charged particle densities as a function of $\eta^*$ for \mbox{$0<p_T^* < 1$ GeV} with relative statistical (stat.) and systematic (sys.) uncertainties given in per cent. The phase space is defined in table 1.}
    \label{tab:pt-curr}
  \end{center}
\end{table}


\renewcommand{\arraystretch}{1.15} 
\begin{table}[h]
  \begin{center}
   \tabcolsep 5.8pt
    \normalsize
    \begin{tabular}{|c|ccc|}
      \hline
     \multicolumn{4}{|c|} {\bf \boldmath 
     $1< p_T^* <$ 10 GeV }  \\
     \hline
     \hline
   \multicolumn{1}{|c|}  {$\eta^*$ range} & $1\!/\!N \cdot {\rm d}n/\rm{d}\eta^*\,$ & stat. & sys.  \\  
    \multicolumn{1}{|c|}  {} & \multicolumn{1}{c}{} &\multicolumn{1}{c}{($\%$)} & \multicolumn{1}{c|}{($\%$)} \\ 
      \hline

                                                   $0.0 - 0.5$  & $0.0807$   & $0.26$   &  $6.9$       \\ 
                                                   $0.5 - 1.0$  & $0.1448$   &$ 0.14$   &  $4.6$       \\ 
                                                   $1.0 - 1.5$  & $0.1835$   & $0.11$   &  $2.4$       \\ 
                                                   $1.5 - 2.0$  & $0.2066$   & $0.10$   &  $1.4$       \\ 
                                                   $2.0 - 2.5$  & $0.2255$   & $0.09$   &  $1.8$       \\   
                                                   $2.5 - 3.0$  & $0.2251$   & $0.09$   &  $1.9$       \\ 
                                                   $3.0 - 3.7$  & $0.1668$   & $0.13$   &  $2.3$       \\  
                                                   $3.7 - 5.0$  & $0.0329$  & $0.42$   &  $4.5$       \\  
      \hline
    \end{tabular}
    \caption{Charged particle densities as a function of $\eta^*$ for \mbox{$1 < p_T^* < 10$ GeV} with relative statistical (stat.) and systematic (sys.) uncertainties given in per cent. The phase space is defined in table 1.}
    \label{tab:pt-curr1}
  \end{center}
\end{table}


\renewcommand{\arraystretch}{1.15} 
\begin{table}[p]
  \begin{center}
   \tabcolsep 5.8pt
\begin{tabular}{ccc}

    \begin{tabular}{|c|c|ccc|}
      \hline
     \multicolumn{5}{|c|} { {\bf \boldmath 
     $0<p_T^* <$ 1 GeV }}  \\
     \hline
     \hline
  \multicolumn{1}{|c|}{ $Q^2$, $x$ intervals} &  \multicolumn{1}{|c|}  {$\eta^*$ range} & $1\!/\!N \cdot {\rm d}n/\rm{d}\eta^*\,$ & stat. & sys.  \\  
   \multicolumn{1}{|c|}{} & \multicolumn{1}{|c|}  {} & \multicolumn{1}{c}{} &\multicolumn{1}{c}{($\%$)} & \multicolumn{1}{c|}{($\%$)} \\ 
      \hline
     \multirow{4}{*}{$5< Q^2 < 10 $ GeV$^{2}$}      & $0.0 - 1.0$  &  $1.398$  & $0.14$   & $5.8$        \\ 
                                                    & $1.0 - 1.6$  &  $1.621$  & $0.09$   & $3.5$        \\ 
      \multirow{3}{*}{$0.0001 <x < 0.00024$}        & $1.6 - 2.1$  &  $1.727$  & $0.08$   & $2.8$        \\ 
                                                    & $2.1 - 2.6$  &  $1.760$  & $0.08$   & $3.5$        \\ 
                                                    & $2.6 - 3.1$  &  $1.749$  & $0.08$   & $3.5$        \\   
                                                    & $3.1 - 3.7$  &  $1.650$  & $0.10$   & $2.6$        \\ 
                                                    & $3.7 - 5.0$  &  $0.683$  & $0.23$   & $2.1$        \\  
      \hline
    \multirow{3}{*}{$5< Q^2 < 10 $ GeV$^{2}$}      &  $0.0 - 1.5$ & $1.241$   & $0.22$   &  $4.4$       \\ 
                                                    &  $1.5 - 2.3$ & $1.682$   & $0.12$   &  $2.7$       \\ 
      \multirow{3}{*}{$0.00024 <x < 0.0005$}        &  $2.3 - 2.8$ & $1.732$   & $0.10$   & $ 3.2$       \\ 
                                                    &  $2.8 - 3.3$ & $1.671$   & $0.10$   &  $2.1$       \\ 
                                                    &  $3.3 - 3.9$ & $1.347$   & $0.12$   &  $2.0$       \\ 
                                                    &  $3.9 - 5.0$ & $0.652$   & $0.24$   &  $2.1$       \\     
      \hline
     \multirow{2}{*}{$5< Q^2 < 10 $ GeV$^{2}$}     &$ 0.5 - 2.0$  & $1.288$  & $0.21$   & $4.4$        \\ 
                                                   & $2.0 - 2.9$  & $1.613$  & $0.13$   & $1.8$       \\ 
      \multirow{2}{*}{$0.0005 <x < 0.002$}         & $2.9 - 3.7$  & $1.272$  & $0.15$   & $2.0$        \\ 
                                                   & $3.7 - 5.0$  & $0.554$  & $0.28$   &$ 2.5$        \\  
      \hline
     \multirow{4}{*}{$10< Q^2 < 20 $ GeV$^{2}$}     & $0.0 - 1.0$  & $1.464$   & $0.15$   &  $5.9$       \\ 
                                                    & $1.0 - 1.6$  & $1.721$   & $0.10$   &  $2.5$       \\ 
      \multirow{3}{*}{$0.0002 <x < 0.00052$}        & $1.6 - 2.1$  & $1.820$   & $0.09$   &  $2.6$      \\ 
                                                    & $2.1 - 2.6$  & $1.865$   & $0.09$   &  $2.8$       \\ 
                                                    & $2.6 - 3.1$  & $1.857$   & $0.09$   &  $2.1$       \\ 
                                                    &$ 3.1 - 3.7$  & $1.680$   & $0.10$   &  $1.6$       \\ 
                                                    & $3.7 - 5.0$  & $0.784$   & $0.23$   &  $1.7$       \\ 
       \hline

    \end{tabular}

\end{tabular}
\quad
    \caption{Charged particle densities as a function of $\eta^*$ for \mbox{$0<p_T^* < 1$ GeV} for different $Q^2$ and $x$ intervals with relative statistical (stat.) and systematic (sys.) uncertainties given in per cent. The phase space is defined in table 1.}
    \label{tab:pt-curr2}
  \end{center}
\end{table}
\begin{table}
\centering
    \begin{tabular}{|c|c|ccc|}
      
      \hline
  \multicolumn{1}{|c|}{ $Q^2$, $x$ intervals} &  \multicolumn{1}{|c|}  {$\eta^*$ range} & $1\!/\!N \cdot {\rm d}n/{\rm d}\eta^*\,$ & stat. & sys.  \\  
   \multicolumn{1}{|c|}{} & \multicolumn{1}{|c|}  {} & \multicolumn{1}{c}{} &\multicolumn{1}{c}{($\%$)} & \multicolumn{1}{c|}{($\%$)} \\ 
      \hline
     \multirow{4}{*}{$10< Q^2 < 20 $ GeV$^{2}$}     & $0.0 - 1.5$  & $1.268$   & $0.22$   &  $5.2$       \\ 
                                                    & $1.5 - 2.3$  & $1.790$   & $0.12$   &  $2.6$       \\ 
      \multirow{3}{*}{$0.00052 <x < 0.0011$}        & $2.3 - 2.8$  & $1.819$   & $0.10$   &  $2.1$       \\ 
                                                    & $2.8 - 3.3$  & $1.685$   & $0.10$   &  $1.7$       \\ 
                                                    & $3.3 - 3.9$  & $1.312$   & $0.12$   &  $1.9$       \\ 
                                                    & $3.9 - 5.0$  & $0.700$   & $0.23$   &  $2.0$       \\  
     \hline
     \multirow{2}{*}{$ 10< Q^2 < 20 $ GeV$^{2}$}    & $0.5 - 2.0$  &$ 1.34$   & $0.20$   & $8.6$        \\ 
                                                    & $2.0 - 2.9$  & $1.650$   & $0.13$   & $1.9$        \\ 
      \multirow{2}{*}{$ 0.0011 <x < 0.0037$}        & $2.9 - 3.7$  & $1.263$   & $0.14$   & $2.3$        \\ 
                                                    & $3.7 - 5.0$  &$ 0.550$   & $0.28$   & $3.4$        \\  
        \hline

     \multirow{4}{*}{$20< Q^2 < 100 $ GeV$^{2}$}    & $0.0 - 1.0$  & $1.461$   & $0.15$   &  $5.8$        \\ 
                                                    & $1.0 - 1.6$  & $1.820$   & $0.09$   &  $2.7$       \\ 
      \multirow{3}{*}{$0.0004 <x < 0.0017$}         & $1.6 - 2.1$  & $1.928$   & $0.08$   &  $2.4$       \\ 
                                                    & $2.1 - 2.6$  & $1.951$   & $0.08$   &  $2.4$       \\ 
                                                    &$ 2.6 - 3.1$  & $1.883$   & $0.08$   &  $2.2$       \\   
                                                    & $3.1 - 3.7$  & $1.601$   & $0.10$   &  $1.7$       \\ 
                                                    & $3.7 - 5.0$  & $0.883$   & $0.21$   &  $2.1$       \\  
      \hline
     \multirow{3}{*}{$20< Q^2 < 100 $ GeV$^{2}$}   &$ 0.0 - 1.5$  & $1.077$   & $0.20$   &  $5.4$       \\ 
                                                   & $1.5 - 2.2$  & $1.783$   & $0.09$   &  $2.6$       \\ 
      \multirow{2}{*}{$0.0017 <x < 0.01$}          & $2.2 - 2.9$  & $1.714$   & $0.09$   &  $1.8$       \\ 
                                                   & $2.9 - 3.7$  & $1.445$   & $0.11$   &  $1.9$       \\ 
                                                   & $3.7 - 5.0$  & $0.634$   & $0.22$   &  $2.0 $      \\   
      \hline

    \end{tabular}
    \captcont{continued}
\end{table}


\renewcommand{\arraystretch}{1.15} 
\begin{table}[p]
 \centering
   \tabcolsep 5.8pt
\begin{tabular}{ccc}

    \begin{tabular}{|c|c|ccc|}
      \hline
     \multicolumn{5}{|c|} {  {\bf \boldmath 
     $1< p_T^* <$ 10 GeV }}  \\
     \hline
     \hline
  \multicolumn{1}{|c|}{ $Q^2$, $x$ intervals} &  \multicolumn{1}{|c|}  {$\eta^*$ range} & $1\!/\!N \cdot {\rm d}n/\rm{d}\eta^*\,$ & stat. & sys.  \\  
   \multicolumn{1}{|c|}{} & \multicolumn{1}{|c|}  {} & \multicolumn{1}{c}{} &\multicolumn{1}{c}{($\%$)} & \multicolumn{1}{c|}{($\%$)} \\ 
      \hline
      \multirow{4}{*}{$5< Q^2 < 10 $ GeV$^{2}$}    & $0.0 - 0.5$  & $0.1365$   &$ 0.39$   & $5.7$        \\ 
                                                   & $0.5 - 1.0$  & $0.1551$   & $0.29$   & $3.5$        \\ 
      \multirow{4}{*}{$0.0001 <x < 0.00024$}       & $1.0 - 1.5$  & $0.1679$   & $0.27$   & $2.3$        \\ 
                                                   & $1.5 - 2.0$  & $0.1818$   & $0.26$   & $3.5$        \\ 
                                                   & $2.0 - 2.5$  & $0.1961$   & $0.26$   & $2.4$        \\   
                                                   & $2.5 - 3.0$  & $0.2084$   &$ 0.24$   & $2.4$        \\ 
                                                   & $3.0 - 3.7$  &$ 0.1984$   & $0.31$   & $2.3$        \\  
                                                   & $3.7 - 5.0$  & $0.0501$   & $1.02$   & $3.6$        \\  
      \hline
     \multirow{4}{*}{$5< Q^2 < 10 $ GeV$^{2}$}      & $0.0 - 1.0$  & $0.1115$   & $0.70$   & $ 7.3$       \\ 
                                                    & $1.0 - 1.5$  & $0.1575$   & $0.33$   &  $3.4$       \\ 
      \multirow{3}{*}{$0.00024 <x < 0.0005$}        & $1.5 - 2.0$  & $0.1758$   & $0.31$   & $ 3.3$       \\ 
                                                    & $2.0 - 2.5$  & $0.1903$   & $0.30$   &  $5.2$       \\ 
                                                    & $2.5 - 3.0$  & $0.2037$   & $0.29$   &  $1.9$       \\  
                                                    & $3.0 - 3.7$  & $0.1626$   & $0.38$   &  $2.6$       \\ 
                                                    & $3.7 - 5.0$  & $0.0311$   & $1.33$   &  $8.0$       \\ 
        \hline
     \multirow{3}{*}{$5< Q^2 < 10 $ GeV$^{2}$}   & $0.5 - 1.5$  & $0.1244$   & $0.66$   & $ 4.5$       \\ 
                                                 & $1.5 - 2.0$  &$ 0.1675$   & $0.32$   &  $4.7$       \\ 
      \multirow{3}{*}{$0.0005 <x < 0.002$}       & $2.0 - 2.5$  & $0.1811$   &$ 0.31$   & $ 3.0$       \\ 
                                                 & $2.5 - 3.0$  &$ 0.1686$   & $0.32$   &  $2.3$       \\ 
                                                 & $3.0 - 3.7$  & $0.0978$   & $0.51$   &  $3.0$       \\   
                                                 & $3.7 - 5.0$  & $0.01071$   & $2.21$   &  $4.7$       \\  
      \hline
      \multirow{4}{*}{$10< Q^2 < 20 $ GeV$^{2}$}   & $0.0 - 0.5$  & $0.1506$   & $0.42$   & $6.1$        \\ 
                                                   & $0.5 - 1.0$  & $0.1764$   & $0.30$   & $2.1$        \\ 
      \multirow{4}{*}{$0.0002 <x < 0.00052$}       & $1.0 - 1.5$  & $0.1959$   & $0.28$   & $2.7$        \\ 
                                                   & $1.5 - 2.0$  & $0.2180$   & $0.26$   & $1.7$        \\ 
                                                   & $2.0 - 2.5$  & $0.2444$   & $0.25$   & $1.9$        \\   
                                                   & $2.5 - 3.0$  & $0.2646$   & $0.23$   & $2.7$        \\ 
                                                   & $3.0 - 3.7$  & $0.2334$   & $0.31$   & $1.4$        \\  
                                                   & $3.7 - 5.0$  & $0.0552$   & $1.03$   & $2.7$       \\  
      \hline
    \end{tabular}

\end{tabular}
    \caption{Charged particle densities as a function of $\eta^*$ for \mbox{$1 < p_T^* < 10$ GeV} for  different $Q^2$ and $x$ intervals with relative statistical (stat.) and systematic (sys.) uncertainties given in per cent. The phase space is defined in table 1.}
    \label{tab:pt-curr3}
\end{table}

\begin{table}
\centering
    \begin{tabular}{|c|c|ccc|}
      
      \hline
  \multicolumn{1}{|c|}{ $Q^2$, $x$ intervals} &  \multicolumn{1}{|c|}  {$\eta^*$ range} & $1\!/\!N \cdot {\rm d}n/\rm{d}\eta^*\,$ & stat. & sys.  \\  
   \multicolumn{1}{|c|}{} & \multicolumn{1}{|c|}  {} & \multicolumn{1}{c}{} &\multicolumn{1}{c}{($\%$)} & \multicolumn{1}{c|}{($\%$)} \\ 
      \hline
     \multirow{4}{*}{$10< Q^2 < 20 $ GeV$^{2}$}     & $0.0 - 1.0$  & $0.123$    & $0.69$   & $8.3$       \\ 
                                                    & $1.0 - 1.5$  & $0.1910$   & $0.30$   & $3.4$        \\ 
      \multirow{3}{*}{$0.00052 <x < 0.0011$}        & $1.5 - 2.0$  &$ 0.2145$   & $0.27$   & $3.1$        \\ 
                                                    & $2.0 - 2.5$  & $0.2409$   & $0.26$   &$ 3.6$        \\ 
                                                    & $2.5 - 3.0$  & $0.2415$   & $0.26$   & $2.6$        \\  
                                                    & $3.0 - 3.7$  & $0.1750$   & $0.37$   & $2.4$        \\ 
                                                    & $3.7 - 5.0$  & $0.0315$   & $1.34$   & $3.7$        \\ 
      \hline
     \multirow{3}{*}{$10< Q^2 < 20 $ GeV$^{2}$}   & $0.5 - 1.5$  & $0.1496$   & $0.58$   & $6.6$        \\ 
                                                  & $1.5 - 2.0$  & $0.2086$   & $0.27$   & $3.8$       \\ 
      \multirow{3}{*}{$0.0011 <x < 0.0037$}       & $2.0 - 2.5$  & $0.2246$   & $0.26$   & $4.5$       \\ 
                                                  & $2.5 - 3.0$  & $0.1997$   & $0.28$   & $3.8$        \\ 
                                                  & $3.0 - 3.7$  & $0.1051$   & $0.48$   & $3.8$        \\   
                                                  &$ 3.7 - 5.0$  & $0.01028$   & $2.38$   &$ 7.7$        \\  
       \hline
      \multirow{4}{*}{$20< Q^2 < 100 $ GeV$^{2}$}  & $0.0 - 0.5$  & $0.1581$   & $0.42$   & $6.1$        \\ 
                                                   & $0.5 - 1.0$  & $0.2105$   & $0.27$   & $2.3$        \\ 
      \multirow{4}{*}{$0.0004 <x < 0.0017$}        &$ 1.0 - 1.5$  & $0.2479$   & $0.23$   &$ 2.5$        \\ 
                                                   & $1.5 - 2.0$  & $0.2820$   & $0.22$   & $2.1$        \\ 
                                                   & $2.0 - 2.5$  & $0.3188$   & $0.20$   & $2.3$        \\   
                                                   & $2.5 - 3.0$  & $0.3386$   & $0.20$   & $1.4$        \\ 
                                                   &$ 3.0 - 3.7$  & $0.2601$   & $0.28$   & $1.5$        \\  
                                                   & $3.7 - 5.0$  & $0.0602$   & $0.95$   & $1.8$        \\  
      \hline
     \multirow{4}{*}{$20< Q^2 < 100 $ GeV$^{2}$}    & $0.0 - 1.0$  & $0.118$    & $0.63$   & $9.0$        \\ 
                                                    & $1.0 - 1.5$  & $0.251$    & $0.22$   & $4.5$        \\ 
      \multirow{3}{*}{$0.0017 <x < 0.01$}           & $1.5 - 2.0$  & $0.2966$   & $0.19$   & $3.4$       \\ 
                                                    & $2.0 - 2.5$  & $0.3167$   & $0.19$   & $2.1$        \\ 
                                                    & $2.5 - 3.0$  & $0.2799$   & $0.20$   & $2.1$        \\  
                                                    & $3.0 - 3.7$  & $0.1587$   & $0.34$   & $2.4$        \\ 
                                                    & $3.7 - 5.0$  & $0.02112$  & $1.49$   & $4.6$        \\ 
      \hline

\end{tabular}
    \captcont{continued}
\end{table}


\newpage

\renewcommand{\arraystretch}{1.15} 
\begin{table}[t]
\vspace{+1.5cm}
  \begin{center}
   \tabcolsep 5.8pt
    \normalsize
    \begin{tabular}{|c|lcc|}
      \hline
     \multicolumn{4}{|c|} {\bf \boldmath $0<\eta^* <1.5$  }  \\
     \hline
     \hline
   \multicolumn{1}{|c|}  {$p_T^*$ range} & $1\!/\!N \cdot {\rm d}n/{\rm d}p_T^*\,$ & stat. & sys.  \\  
    \multicolumn{1}{|c|}{[GeV]} & \multicolumn{1}{c}{[GeV$^{-1}$]} &\multicolumn{1}{c}{($\%$)} & \multicolumn{1}{c|}{($\%$)} \\ 
      \hline
                                                 $ 0.2 - 0.4$  & $\ 3.952$  & $0.01$   & $2.0$        \\ 
                                                  $ 0.4 - 0.6$  & $\ 2.431$  & $0.02$   & $1.6$        \\ 
                                                  $ 0.6 - 1.0$  & $\ 0.954$  & $0.04$   & $1.8$        \\ 
                                                   $1.0 - 2.0$  & $\ 0.1686$  & $ 0.15$  & $2.5$       \\ 
                                                   $2.0 - 4.0$  & $\ 1.549\cdot 10^{-2}$  & $0.70$   & $2.0$        \\   
                                                   $4.0 - 10.0$ & $\ 7.15\cdot 10^{-4}$  & $5.4$1   & $1.9$       \\  
      \hline
    \end{tabular}
    \caption{Charged particle densities as a function as a function of $p_T^*$ in the region \mbox{$0<\eta^* <1.5$}  shown with relative statistical (stat.) and systematic (sys.) uncertainties given in per cent. The phase space is defined in table 1.}
    \label{tab:pt-curr4}
  \end{center}
\end{table}

\renewcommand{\arraystretch}{1.15} 
\begin{table}[h]
  \begin{center}
   \tabcolsep 5.8pt
    \normalsize
    \begin{tabular}{|c|lcc|}
      \hline
     \multicolumn{4}{|c|} {\bf \boldmath $1.5<\eta^* <5$  }  \\
     \hline
     \hline
   \multicolumn{1}{|c|}  {$p_T^*$ range} & $1\!/\!N \cdot {\rm d}n/{\rm d}p_T^*\,$ & stat. & sys.  \\  
    \multicolumn{1}{|c|}{[GeV]} & \multicolumn{1}{c}{[GeV$^{-1}$]} &\multicolumn{1}{c}{($\%$)} & \multicolumn{1}{c|}{($\%$)} \\ 
      \hline
                                                   $0.0 - 0.3$  & $\ 5.24$   & $0.01$   & $1.8$        \\ 
                                                   $0.3 - 0.6$  & $\ 6.10$   & $0.01$   & $1.7$         \\ 
                                                   $0.6 - 1.0$  & $\ 2.234$   & $0.02$   & $1.8$       \\ 
                                                   $1.0 - 1.5$  & $\ 0.6193$   & $0.05$   & $1.5$        \\ 
                                                   $1.5 - 2.1$  & $\ 0.1849$   & $0.10$   & $1.5$        \\   
                                                   $2.1 - 3.0$  & $\ 5.23\cdot 10 ^{-2}$   & $0.23$   & $2.0$        \\ 
                                                   $3.0 - 4.0$  & $\ 1.381\cdot 10 ^{-2}$   & $0.47$   & $2.0$        \\  
                                                   $4.0 - 5.0$  & $\ 4.14\cdot 10 ^{-3}$   & $0.84$   & $2.4$        \\ 
                                                   $5.0 - 6.3$  & $\ 1.402\cdot 10 ^{-3}$   & $1.67$   & $2.8$        \\   
                                                   $6.3 - 7.9$  & $\ 3.98\cdot 10 ^{-4}$   & $3.47$   & $2.5$        \\ 
                                                   $7.9 - 10.0$ & $\ 1.061\cdot 10 ^{-4}$   & $7.60$   & $3.2$        \\   
      \hline
    \end{tabular}
    \caption{Charged particle densities as a function as a function of $p_T^*$ in the region \mbox{$1.5<\eta^* <5$}  shown with relative statistical (stat.) and systematic (sys.) uncertainties given in per cent. The phase space is defined in table 1.}
    \label{tab:pt-curr5}
  \end{center}
\end{table}


\renewcommand{\arraystretch}{1.15} 
\begin{table}[tb]
  \centering
   \tabcolsep 5.8pt
    \begin{tabular}{|c|c|lcc|}
      \hline
     \multicolumn{5}{|c|} { {\bf \boldmath $0 <\eta^* <1.5$ } }  \\
     \hline
     \hline
  \multicolumn{1}{|c|}{ $Q^2$, $x$ intervals} &  \multicolumn{1}{|c|}  {$p_T^*$ range} & $1\!/\!N \cdot {\rm d}n/{\rm d}p_T^*\,$ & stat. & sys.  \\  
   \multicolumn{1}{|c|}{} & \multicolumn{1}{|c|}{[GeV]} & \multicolumn{1}{c}{[GeV$^{-1}$]} &\multicolumn{1}{c}{($\%$)} & \multicolumn{1}{c|}{($\%$)} \\ 
      \hline
      \multirow{3}{*}{$5< Q^2 < 10 $ GeV$^{2}$}    & $0.2 - 0.4$  & $\ 4.76$  & $0.03$   & $3.2$        \\ 
                                                   & $0.4 - 0.6$  & $\ 2.92$  & $0.04$   & $2.9$        \\ 
      \multirow{3}{*}{$0.0001 <x < 0.00024$}       & $0.6 - 1.0$  &$ \ 1.144$  & $0.10$   & $3.7$        \\ 
                                                   & $1.0 - 2.0$  & $\ 0.1955$  & $0.39$   & $3.0$        \\ 
                                                   & $2.0 - 4.0$  & $\ 1.489\cdot10^{-2}$ & $1.89$   & $3.2$        \\   
                                                   & $4.0 - 10.0$ & $\ 5.69\cdot 10^{-4}$ & $15.36$  & $6.0$       \\ 
      \hline
      \multirow{3}{*}{$5< Q^2 < 10 $ GeV$^{2}$}    & $0.2 - 0.4$  & $\ 3.99$  & $0.05$   & $2.5$        \\ 
                                                   & $0.4 - 0.6$  & $\ 2.53$  & $0.06$   & $2.6$        \\ 
      \multirow{3}{*}{$0.00024 <x < 0.0005$}       & $0.6 - 1.0$  & $\ 0.994$  & $0.14$   &$ 2.9$        \\ 
                                                   & $1.0 - 2.0$  & $\ 0.1611$  & $0.52$   & $3.3$        \\ 
                                                   & $2.0 - 4.0$  &$ \ 1.286\cdot 10 ^{-2}$  & $2.58$   &$ 2.7$     \\   
                                                   & $4.0 - 10.0$ & $\ 5.1\cdot 10 ^{-4}$  & $21.40$  & $4.9$     \\ 
      \hline
    \multirow{3}{*}{$5< Q^2 < 10 $ GeV$^{2}$}     & $0.2 - 0.6$  & $\ 2.097$   & $0.11$   & $3.0$        \\ 
                                                   & $0.6 - 1.0$  & $\ 0.659$   & $0.19$   & $3.4$        \\ 
      \multirow{2}{*}{$0.0005 <x < 0.002$}         & $1.0 - 2.0$  & $\ 0.1113$   & $0.74$   & $2.8 $       \\ 
                                                   & $2.0 - 4.0$  & $\ 8.82\cdot 10^{-3}$   & $3.40$   & $3.1$        \\  
                                                   & $4.0 - 10.0$ & $\ 3.3\cdot 10^{-4}$   & $34.47$  & $5.5$        \\          
      \hline
     \end{tabular}
    \caption{Charged particle densities as a function of $p_T^*$ in the region \mbox{$0<\eta^* <1.5$} for different $Q^2$ and $x$ intervals shown with relative statistical (stat.) and systematic (sys.) uncertainties given in per cent. The phase space is defined in table 1.}
    \label{tab:pt-curr6}
\end{table}
\begin{table}
\centering
    \begin{tabular}{|c|c|lcc|}
      
      \hline
  \multicolumn{1}{|c|}{ $Q^2$, $x$ intervals} &  \multicolumn{1}{|c|}  {$p_T^*$ range} & $1\!/\!N \cdot {\rm d}n/{\rm d}p_T^*\,$ & stat. & sys.  \\  
   \multicolumn{1}{|c|}{} & \multicolumn{1}{|c|}{[GeV]} & \multicolumn{1}{c}{[GeV$^{-1}$]} &\multicolumn{1}{c}{($\%$)} & \multicolumn{1}{c|}{($\%$)} \\ 
      \hline
      \multirow{3}{*}{$10< Q^2 < 20 $ GeV$^{2}$}   & $0.2 - 0.4$  & $\ 4.97$  & $0.04$   & $2.5$        \\ 
                                                   & $0.4 - 0.6$  & $\ 3.060$  & $0.05$   & $2.7$        \\ 
      \multirow{3}{*}{$0.0002 <x < 0.00052$}       & $0.6 - 1.0$  & $\ 1.229$  & $0.11$   & $2.6$        \\ 
                                                   & $1.0 - 2.0$  & $\ 0.2155$  & $0.40 $  & $3.5$        \\ 
                                                   & $2.0 - 4.0$  & $\ 1.960\cdot 10^{-2}$  & $1.87$   & $3.1 $       \\   
                                                   & $4.0 - 10.0$ & $\ 9.2\cdot 10^{-4}$  & $14.13$  & $3.3$       \\ 
      \hline
      \multirow{3}{*}{$10< Q^2 < 20 $ GeV$^{2}$}   & $0.2 - 0.4$  & $\ 4.05$  & $0.05$   & $3.0$        \\ 
                                                   & $0.4 - 0.6$  & $\ 2.593$  & $0.06$   & $2.9$        \\ 
      \multirow{3}{*}{$0.00052 <x < 0.0011$}       &$ 0.6 - 1.0$  & $\ 1.033$  & $0.13$   & $3.5$        \\ 
                                                   &$ 1.0 - 2.0$  & $\ 0.1811$  & $0.50$   & $4.2$        \\ 
                                                   & $2.0 - 4.0$  & $\ 1.623 \cdot 10 ^{-2}$  & $2.34$   & $5.8$        \\   
                                                   & $4.0 - 10.0$ & $\ 7.5 \cdot 10 ^{-4}$  & $18.91$  & $3.2$       \\ 
      \hline
     \multirow{3}{*}{$10< Q^2 < 20 $ GeV$^{2}$}    & $0.2 - 0.6$  &$ \ 2.124$   & $0.10$   & $4.5$        \\ 
                                                   & $0.6 - 1.0$  &$ \ 0.692$   & $0.18$   &$ 4.1 $       \\ 
      \multirow{2}{*}{$0.0011 <x < 0.0037$}        &$ 1.0 - 2.0$  &$ \ 0.129$   &$ 0.66$   & $4.8$        \\ 
                                                   & $2.0 - 4.0$  &$ \ 1.119 \cdot 10 ^{-2}$   & $3.12$   & $3.0$        \\  
                                                   & $4.0 - 10.0$ & $\ 4.42 \cdot 10 ^{-4}$   & $28.92$  &  $6.0$        \\          
      \hline
      \multirow{3}{*}{$20< Q^2 < 100 $ GeV$^{2}$}  & $0.2 - 0.4$  & $\ 5.00$  & $0.04$   & $2.5$        \\ 
                                                   & $0.4 - 0.6$  & $\ 3.156$  & $0.04$   &$ 2.4$        \\ 
      \multirow{3}{*}{$0.0004 <x < 0.0017$}        & $0.6 - 1.0$  & $\ 1.296$  & $0.97$   & $2.8$        \\ 
                                                   & $1.0 - 2.0$  & $\ 0.2474$  &$ 0.36$   & $3.7$        \\ 
                                                   & $2.0 - 4.0$  & $\ 2.579 \cdot 10 ^{-2}$  & $1.57$   &$ 4.7$     \\   
                                                   & $4.0 - 10.0$ & $\ 1.40 \cdot 10 ^{-3}$  & $11.28$  & $2.8$     \\ 
      \hline
      \multirow{3}{*}{$20< Q^2 < 100 $ GeV$^{2}$}  & $0.2 - 0.4$  & $\ 3.356$  & $0.05$   &  $2.8$       \\ 
                                                   & $0.4 - 0.6$  & $\ 2.254$  & $0.05$   &  $2.8$      \\ 
      \multirow{3}{*}{$0.0017 <x < 0.01$}          & $0.6 - 1.0$  & $\ 0.945$  & $0.12$   &  $3.6$       \\ 
                                                   & $1.0 - 2.0$  & $\ 0.1905$  & $0.42$   &  $5.4$       \\ 
                                                   & $2.0 - 4.0$  & $\ 2.185 \cdot 10 ^{-2}$   & $1.74$   &$ 6.9 $       \\   
                                                   & $4.0 - 10.0$ & $\ 1.21 \cdot 10 ^{-3}$   & $1 3.18$  & $3.5$       \\ 
      \hline
    \end{tabular}
    \captcont{continued}
\end{table}



\renewcommand{\arraystretch}{1.15} 
\begin{table}[p]
\centering
   \tabcolsep 5.8pt

    \begin{tabular}{|c|c|lcc|}
      \hline
     \multicolumn{5}{|c|} { {\bf \boldmath $1.5<\eta^* <5$ } }  \\
     \hline
     \hline
  \multicolumn{1}{|c|}{ $Q^2$, $x$ intervals} &  \multicolumn{1}{|c|}  {$p_T^*$ range} & $1\!/\!N \cdot {\rm d}n/{\rm d}p_T^*\,$ & stat. & sys.  \\  
   \multicolumn{1}{|c|}{} & \multicolumn{1}{|c|}{[GeV]} & \multicolumn{1}{c}{[GeV$^{-1}$]} &\multicolumn{1}{c}{($\%$)} & \multicolumn{1}{c|}{($\%$)} \\ 
      \hline
      \multirow{6}{*}{$5< Q^2 < 10 $ GeV$^{2}$}    & $0.0 - 0.3$   &$\  5.60   $&$ 0.04   $&$ 2.5$        \\ 
                                                   & $0.3 - 0.6$   &$\  6.83   $&$ 0.04   $&$ 2.8$         \\ 
      \multirow{5}{*}{$0.0001 <x < 0.00024$}       & $0.6 - 1.0$   &$ \ 2.290   $&$ 0.07   $&$ 3.6$        \\ 
                                                   & $1.0 - 1.5$   &$ \ 0.639   $&$ 0.14   $&$ 3.3$        \\ 
                                                   & $1.5 - 2.1$   &$ \ 0.1897   $&$ 0.30   $&$ 2.7$        \\   
                                                   & $2.1 - 3.0$   &$ \ 5.12 \cdot 10 ^{-2}    $&$ 0.69   $&$ 4.0$         \\ 
                                                   & $3.0 - 4.0$   &$ \ 1.378 \cdot 10 ^{-2}    $&$ 1.40   $&$ 4.8 $       \\  
                                                   & $4.0 - 5.0$   &$ \ 3.97 \cdot 10 ^{-3}    $&$ 2.46   $&$ 8.6$        \\ 
                                                   & $5.0 - 6.3$   &$ \ 1.40 \cdot 10 ^{-3}    $&$ 4.68   $&$ 9.7$        \\   
                                                   &$ 6.3 - 7.9$   &$ \ 4.34 \cdot 10 ^{-4}    $&$ 9.33   $&$ 6.2$        \\ 
                                                   & $7.9 - 10.0$  &$ \ 9.3 \cdot 10 ^{-5}    $&$ 20.80  $&$ 13.2$        \\   
      \hline
      \multirow{6}{*}{$5< Q^2 < 10 $ GeV$^{2}$}    & $0.0 - 0.3$   &$ \ 5.57   $&$ 0.04   $&$ 2.8 $       \\ 
                                                   & $0.3 - 0.6$   &$ \ 6.64   $&$ 0.04   $&$ 2.6 $       \\ 
      \multirow{5}{*}{$0.00024 <x < 0.0005$}       & $0.6 - 1.0$   &$ \ 2.207   $&$ 0.08   $&$ 2.2 $       \\ 
                                                   & $1.0 - 1.5$   &$ \ 0.587   $&$ 0.17   $&$ 1.7 $       \\ 
                                                   & $1.5 - 2.1$   &$ \ 0.161   $&$ 0.36   $&$ 1.8 $       \\   
                                                   & $2.1 - 3.0$   &$ \ 4.12 \cdot 10 ^{-2}    $&$ 0.85   $&$ 3.2  $      \\ 
                                                   & $3.0 - 4.0$   &$ \ 1.002 \cdot 10 ^{-2}    $&$ 1.83   $&$ 5.7 $       \\  
                                                   & $4.0 - 5.0$   &$ \ 2.63 \cdot 10 ^{-3}    $&$ 3.46   $&$ 6.7  $      \\ 
                                                   & $5.0 - 6.3$   &$ \ 7.69 \cdot 10 ^{-4}    $&$ 7.08   $&$ 5.7  $      \\   
                                                   &$ 6.3 - 7.9$   &$ \ 2.46 \cdot 10 ^{-4}    $&$ 13.98  $&$ 10.4 $       \\ 
                                                   & $7.9 - 10.0$ &$ \ 6.65 \cdot 10 ^{-5}    $&$ 32.90  $&$ 18.6  $      \\   
      \hline
      \multirow{6}{*}{$5< Q^2 < 10 $ GeV$^{2}$}    & $0.0 - 0.3$  &$ \ 4.96   $&$ 0.05   $&$ 2.2$        \\ 
                                                   &$ 0.3 - 0.6$  &$ \ 5.74   $&$ 0.04   $&$ 2.1$        \\ 
      \multirow{5}{*}{$0.0005 <x < 0.002$}         & $0.6 - 1.0$  &$ \ 1.863   $&$ 0.08   $&$ 3.4$        \\ 
                                                   & $1.0 - 1.5$   &$ \ 0.472   $&$ 0.19   $&$ 2.4$        \\ 
                                                   & $1.5 - 2.1$  &$ \ 0.1214   $&$ 0.40   $&$ 2.5$        \\   
                                                   & $2.1 - 3.0$  &$ \ 2.85 \cdot 10 ^{-2}   $&$ 1.02   $&$ 5.5$      \\ 
                                                   & $3.0 - 4.0$  &$ \ 6.14 \cdot 10 ^{-3}   $&$ 2.40   $&$ 6.1$      \\  
                                                   & $4.0 - 5.0$  &$ \ 1.55 \cdot 10 ^{-3}   $&$ 4.94   $&$ 5.4$      \\ 
                                                   & $5.0 - 6.3$  &$ \ 4.401 \cdot 10 ^{-4}   $&$ 10.49  $&$ 12.8$      \\   
                                                   & $6.3 - 7.9$  &$ \ 1.11 \cdot 10 ^{-4}   $&$ 24.17  $&$ 13.6$       \\   
       \hline
    \end{tabular}
    \caption{Charged particle densities as a function of $p_T^*$ in the region \mbox{$1.5<\eta^* <5$} for different $Q^2$ and $x$ intervals shown with relative statistical (stat.) and systematic (sys.) uncertainties given in per cent. The phase space is defined in table 1.}
    \label{tab:pt-curr7}
\end{table}

\begin{table}
\centering
    \begin{tabular}{|c|c|lcc|}
      
      \hline
  \multicolumn{1}{|c|}{ $Q^2$, $x$ intervals} &  \multicolumn{1}{|c|}  {$p_T^*$ range} & $1\!/\!N \cdot {\rm d}n/{\rm d}p_T^*\,$ & stat. & sys.  \\  
   \multicolumn{1}{|c|}{} & \multicolumn{1}{|c|}{[GeV]} & \multicolumn{1}{c}{[GeV$^{-1}$]} &\multicolumn{1}{c}{($\%$)} & \multicolumn{1}{c|}{($\%$)} \\ 
      \hline
      \multirow{6}{*}{$10< Q^2 < 20 $ GeV$^{2}$}   & $ 0.0 - 0.3$  & $\ 5.975$   & $0.04$   & $1.5$        \\ 
                                                   & $0.3 - 0.6$  & $\ 7.16$   & $0.04$   & $2.0$        \\ 
      \multirow{5}{*}{$0.0002 <x < 0.00052$}       & $0.6 - 1.0$  & $\ 2.517$   &$ 0.07$   & $1.7$        \\ 
                                                   & $1.0 - 1.5$  & $\ 0.742$   & $0.15$   & $2.4$        \\ 
                                                   & $1.5 - 2.1$  & $\ 0.2317$   & $0.29$   & $2.3$        \\   
                                                   & $2.1 - 3.0$  & $\ 6.82 \cdot 10 ^{-2}$   & $0.64$   & $3.4$        \\ 
                                                   & $3.0 - 4.0$  &$ \ 1.848 \cdot 10 ^{-2}$   & $1.30$   & $3.6$        \\  
                                                   & $4.0 - 5.0$  &$ \ 5.83 \cdot 10 ^{-3}$   & $2.27$   &$ 5.2$        \\ 
                                                   & $5.0 - 6.3$  & $\ 2.00 \cdot 10 ^{-3}$   & $4.45$   & $2.5$        \\   
                                                   &$ 6.3 - 7.9$  & $\ 5.88 \cdot 10 ^{-4}$   & $8.91$   & $2.9 $       \\ 
                                                   & $7.9 - 10.0$ & $\ 1.942 \cdot 10 ^{-4}$   &$ 17.77$  & $5.3$        \\   
      \hline
      \multirow{6}{*}{$10< Q^2 < 20 $ GeV$^{2}$}   &$ 0.0 - 0.3$  & $\ 5.824$   & $0.05$   &$ 1.5 $      \\ 
                                                   &$ 0.3 - 0.6$  & $\ 6.77$   & $0.04$   & $1.7 $       \\ 
      \multirow{5}{*}{$0.00052 <x < 0.0011$}       &$ 0.6 - 1.0$  & $\ 2.332$   &$ 0.08$   & $2.2$        \\ 
                                                   &$ 1.0 - 1.5$  & $\ 0.657$   &$ 0.16$   & $1.8 $       \\ 
                                                   &$ 1.5 - 2.1$  &$ \ 0.1948$   &$ 0.31$   &$ 2.3 $       \\   
                                                   &$ 2.1 - 3.0$  &$ \ 0.053 8$   &$ 0.74$   & $3.0 $     \\ 
                                                   &$ 3.0 - 4.0$  & $\ 1.297 \cdot 10 ^{-2}$   & $1.56$   &$ 5.9 $       \\  
                                                   &$ 4.0 - 5.0$  & $\ 3.94 \cdot 10 ^{-3}$   & $2.90$   & $4.5  $      \\ 
                                                   &$ 5.0 - 6.3$  & $\ 1.33 \cdot 10 ^{-3}$   & $5.80$   &$ 6.9 $       \\   
                                                   &$ 6.3 - 7.9$  & $\ 3.40 \cdot 10 ^{-4}$   & $12.91$  &$ 6.3 $       \\ 
                                                   &$ 7.9 - 10.0$ & $\ 7.7 \cdot 10 ^{-5}$   & $30.96$  & $8.6  $      \\   
      \hline
      \multirow{5}{*}{$10< Q^2 < 20 $ GeV$^{2}$}   &$ 0.0 - 0.3$  & $\ 5.12$   &$ 0.05$   &  $2.2$       \\ 
                                                   &$ 0.3 - 0.6$  & $\ 5.65$   & $0.04$   & $ 2.4 $      \\ 
      \multirow{5}{*}{$0.0011 <x < 0.0037$}        &$ 0.6 - 1.0$  & $\ 1.946$   &$ 0.08$   & $ 2.0$       \\ 
                                                   &$ 1.0 - 1.5$  & $\ 0.538 $  & $0.17$   &  $2.2 $      \\ 
                                                   &$ 1.5 - 2.1$  & $\ 0.1481$   & $0.35$   & $ 3.9 $      \\   
                                                   &$ 2.1 - 3.0$  & $\ 3.89 \cdot 10 ^{-2}$   & $0.85 $  & $3.6$        \\ 
                                                   &$ 3.0 - 4.0$  & $\ 8.643 \cdot 10 ^{-3}$   &$ 1.94$   & $3.4$        \\  
                                                   &$ 4.0 - 5.0$  & $\ 2.10 \cdot 10 ^{-3}$   & $3.93 $  & $4.7$        \\ 
                                                   &$ 5.0 - 6.3$  & $\ 6.43 \cdot 10 ^{-4}$   & $8.90$   &$ 12.9$        \\   
                                                   &$ 6.3 - 10.0$ & $\ 7.7 \cdot 10 ^{-5}$   & $46.70$  & $10.6$        \\ 
       \hline

\end{tabular}
    \captcont{continued}
\end{table}

 
 \begin{table}
\centering
    \begin{tabular}{|c|c|lcc|}
      
      \hline
  \multicolumn{1}{|c|}{ $Q^2$, $x$ intervals} &  \multicolumn{1}{|c|}  {$p_T^*$ range} & $1\!/\!N \cdot {\rm d}n/{\rm d}p_T^*\,$ & stat. & sys.  \\  
   \multicolumn{1}{|c|}{} & \multicolumn{1}{|c|}{[GeV]} & \multicolumn{1}{c}{[GeV$^{-1}$]} &\multicolumn{1}{c}{($\%$)} & \multicolumn{1}{c|}{($\%$)} \\ 
      \hline

      \multirow{6}{*}{$20< Q^2 < 100 $ GeV$^{2}$}  & $0.0 - 0.3$  & $\ 6.25 $  & $0.04$   & $1.6 $      \\ 
                                                   & $0.3 - 0.6$  & $\ 7.32 $  & $0.04$   & $1.7 $       \\ 
      \multirow{5}{*}{$0.0004 <x < 0.0017$}        & $0.6 - 1.0$  & $\ 2.673$   & $0.07$   &$ 2.1$        \\ 
                                                   & $1.0 - 1.5$  & $\ 0.851 $  & $0.13$   & $2.5$        \\ 
                                                   & $1.5 - 2.1$  & $\ 0.2898$   & $0.24$   & $2.4 $      \\   
                                                   & $2.1 - 3.0$  & $\ 9.21 \cdot 10 ^{-2}$   & $0.52$   & $2.9$       \\ 
                                                   & $3.0 - 4.0$  &$ \ 2.790 \cdot 10 ^{-2}$   & $1.01$   & $2.2$        \\  
                                                   & $4.0 - 5.0$  & $\ 9.30 \cdot 10 ^{-3}$   &$ 1.75$   & $3.0$       \\ 
                                                   & $5.0 - 6.3$  & $\ 3.28 \cdot 10 ^{-3}$   & $3.37$   & $3.4$        \\   
                                                   & $6.3 - 7.9 $ & $\ 9.64 \cdot 10 ^{-4}$   & $6.79$   & $4.2$        \\ 
                                                   & $7.9 - 10.0$ & $\ 3.06 \cdot 10 ^{-4}$   & $14.41$  & $3.9$       \\   
      \hline
      \multirow{6}{*}{$20< Q^2 < 100 $ GeV$^{2}$}  & $0.0 - 0.3$  &$ \ 5.652 $ & $0.04$   & $1.5 $       \\ 
                                                   & $0.3 - 0.6$  & $\ 6.052 $  &$ 0.04 $ & $1.6 $      \\ 
      \multirow{5}{*}{$0.0017 <x < 0.01$}          & $0.6 - 1.0$  &$ \ 2.244 $ &$ 0.06 $  & $1.7  $      \\ 
                                                   & $1.0 - 1.5$  &$ \ 0.697 $ &$ 0.13$   & $1.8 $       \\ 
                                                   & $1.5 - 2.1$  &$ \ 0.2323$  &$ 0.24$   & $1.9$        \\   
                                                   &$ 2.1 - 3.0$  &$ \ 7.12 \cdot 10 ^{-2}$  & $0.52 $  & $2.1 $    \\ 
                                                   & $3.0 - 4.0$  &$ \ 2.025 \cdot 10 ^{-2}$  & $1.05$   & $1.9  $     \\  
                                                   & $4.0 - 5.0$  &$ \ 6.32 \cdot 10 ^{-3}$  & $1.92$   & $3.9  $      \\ 
                                                   & $5.0 - 6.3$  &$ \ 2.20 \cdot 10 ^{-3}$  & $3.87$   & $4.0  $      \\   
                                                   & $6.3 - 7.9 $ &$ \ 5.91 \cdot 10 ^{-4}$  & $8.42 $  & $6.0$        \\ 
                                                   & $7.9 - 10.0$ &$ \ 1.38 \cdot 10 ^{-4}$  & $19.37 $ & $20.1$        \\   
      \hline

\end{tabular}
    \captcont{continued}
\end{table}


\clearpage


\newpage

\begin{thebibliography}{10}

\bibitem{DGLAP}
V.~Gribov and L.~Lipatov,
Sov.\ J.\ Nucl.\ Phys. {\bf 15} (1972) 438 and 675; \\
L.~Lipatov, 
Sov.\ J.\ Nucl.\ Phys. {\bf 20} (1975) 94; \\
G.~Altarelli and G.~Parisi,
Nucl.\ Phys. {\bf B126} (1977) 298; \\
Y.~Dokshitzer, 
Sov.\ Phys.\ JETP {\bf 46} (1977) 641.

\bibitem{BFKL}
E. Kuraev, L. Lipatov and V. Fadin,  
Sov. Phys. JETP {\bf 44} (1976) 443; \\
E. Kuraev, L. Lipatov and V. Fadin, 
Sov. Phys. JETP {\bf 45} (1977) 199; \\
Y. Balitsky and L. Lipatov,  
Sov. J. Nucl. Phys. {\bf 28} (1978) 822.

\bibitem{CCFM}
M.~Ciafaloni, 
Nucl. Phys. {\bf B296} (1988) 49; \\
M.~Catani, F.~Fiorani and G.~Marchesini,  
Phys.\ Lett. {\bf B234} (1990) 339; \\
S. Catani, F. Fiorani and G. Marchesini,  
Nucl. Phys. {\bf B336} (1990) 18; \\
G. Marchesini, 
Nucl. Phys. {\bf B445} (1995) 49.

\bibitem{F2}
F.D. Aaron {\it et al.}~[H1 and ZEUS Collaborations], JHEP {\bf 1001} (2010) 109 [arXiv:0911.0884];\\
F.D. Aaron {\it et al.}~[H1~Collaboration],
JHEP  {\bf 1209} (2012) 061 [arXiv:1206.7007].


\bibitem{F2-theory-MRST}
R.S. Thorne, Phys. Rev. {\bf D73} (2006) 054019 [hep-ph/0601245]; \\
A.D. Martin {\it et al.}, Eur. Phys. J. {\bf C63} (2009) 189 [arXiv:0901.0002]; \\
R.S. Thorne [arXiv:1201.6180].

\bibitem{F2-theory-CT}
H.L. Lai {\it et al.}, Phys. Rev. {\bf D82}, (2010) 074024 [arXiv:1007.2241]; \\
P. Nadolsky {\it et al.}, contribution to the Proceedings of the XX Workshop on Deep Inelastic
Scattering and Related Subjects, Bonn, Germany, 26-30 March, 2012 [arXiv:1206.3321].

\bibitem{F2-theory-NNPDF}
S. Forte {\it et al.}, Nucl. Phys. {\bf B834}, (2010) 116 [arXiv:1001.2312]; \\
R.D. Ball {\it et al.} [NNPDF Collaboration], Nucl. Phys. {\bf B849} (2011) 296
[arXiv:1101.1300]; \\
R.D. Ball {\it et al.} [NNPDF Collaboration], Nucl. Phys. {\bf B855} (2012) 153
[arXiv:1107.2652].

\bibitem{F2-theory-ABM}
S. Alekhin and S. Moch, Phys. Lett. {\bf B699} (2011) 345
[arXiv:1011.5790]; \\
S. Alekhin {\it et al.}, Phys. Rev. {\bf D81} (2010) 014032 [arXiv:0908.2766]; \\
S. Alekhin and S. Moch [arXiv:1107.0469].


\bibitem{dijet}
A. Aktas {\it et al.}~[H1~Collaboration], 
Eur. Phys. J. {\bf C33} (2004) 477 [hep-ex/0310019].

\bibitem{forw-incl-jet-1}
C.~Adloff {\it et al.}~[H1~Collaboration],
Phys.\ Lett. {\bf B415} (1997) 418
\newblock [hep-ex/9709017].

\bibitem{tr-energy-flow-1}
C.~Adloff {\it et al.}~[H1~Collaboration], 
Eur.\ Phys.\ J. {\bf C12} (2000) 595
[hep-ex/9907027].

\bibitem{tr-energy-flow-2}
S.~Aid {\it et al.}~[H1~Collaboration],
\newblock  Eur.\ Phys. {\bf B356} (1995) 118
\newblock [hep-ex/9506012]. 


\bibitem{forw-jet-h1-zeus1}
A.~Aktas {\it et al.}~[H1~Collaboration],
Eur. Phys. J. {\bf C46} (2006) 27 [hep-ex/0508055].

\bibitem{forw-jet-h1-zeus2}
S.~Chekanov {\it et al.}~[ZEUS~Collaboration],
Eur. Phys. J. {\bf C52} (2007) 515 [arXiv:0707.3093].

\bibitem{forw-jet-h1-zeus3}
F.D. Aaron {\it et al.}~[H1~Collaboration], 
Eur. Phys. J. {\bf C72} (2012) 1910 [arXiv:1111.4227].

\bibitem{fwd-pi}
A.~Aktas {\it et al.}~[H1~Collaboration],
\newblock  Eur. Phys. J. {\bf C36} (2004) 441 [hep-ex/0404009].


\bibitem{Kuhlen-pt}
M.~Kuhlen, 
Phys. Lett. {\bf B382} (1996) 441 [hep-ph/9606246].

\bibitem{pt-old}
C.~Adloff {\it et al.}~[H1~Collaboration],
Nucl.\ Phys {\bf B485} (1997) 3 [hep-ex/9610006].

\bibitem{RAPGAP}
H.~Jung, {\sc Rapgap} 3.1, 
Comput.\ Phys.\ Commun {\bf 86} (1995) 147.

\bibitem{DJANGOH}
K.~Charchula, G.~A.~Schuler and H.~Spiesberger,
\newblock {\sc Djangoh} 1.4,
\newblock Comput.\ Phys.\ Commun. {\bf 81} (1994) 381.

\bibitem{ARIADNE}
L.~L\"onnblad,
\newblock {\sc Ariadne} 4.10,
\newblock Comput.\ Phys.\ Commun. {\bf 71} (1992) 15.

\bibitem{CDM-BFKL}
L.~L\"onnblad, Z.Phys.
{\bf C65} (1995) 285; \\
A.H. Mueller,
Nucl. Phys. {\bf B415} (1994) 373.

\bibitem{CASCADE}
H.~Jung,
Comput.\ Phys.\ Commun. {\bf 143} (2002) 100; \\
H.~Jung {\it et al.}, 
{\sc Cascade} 2.2.0,
Eur.\ Phys.\ J. {\bf C70} (2010) 1237.


\bibitem{herwig}
S.~Gieseke {\it et al.}, Herwig++ 2.5, [arXiv:1102.1672].


\bibitem{POWHEG-DIS}
L. D'Errico and P. Richardson,
Eur. Phys. J. {\bf C72} (2011) 2042 [arXiv:1106.2983].


\bibitem{CoherentPB1}
G.~Marchesini,
\newblock  Nucl. Phys. {\bf B445} (1995) 49
\newblock [hep-ph/9412327].

\bibitem{CoherentPB2}
S.~Catani, B.R.~Webber and G.~Marchesini,
\newblock Nucl. Phys. {\bf B349} (1991) 635.


\bibitem{Phojet}
R. Engel, Z. Phys. {\bf C66}  (1995) 203; \\
R. Engel and J. Ranft, {\sc Phojet} 1.6, Phys. Rev. {\bf D54} (1996) 4244 [hep-ph/9509373].

\bibitem{Phojet1}
A. Capella {\it et al.}, Phys. Rept. {\bf 236} (1994) 227.

\bibitem{cteq6lo}
J.~Pumplin {\it et al.},
JHEP {\bf 0207} (2002) 012 [hep-ph/0201195].

\bibitem{A0}
H. Jung, in ``Proceedings of the XII International Workshop on Deep Inelastic Scattering
(DIS2004)'', eds. D. Bruncko, J. Ferencei, P. Str\'{\i}\v{z}enec, \v{S}trbsk\'{e} Pleso, Slovakia (2004),
299 [hep-ph/0411287].

\bibitem{MRST}
A.D.~Martin {\it et al.},
Phys. Lett. {\bf B531} (2002) 216 [hep-ph/0201127].

\bibitem{Lund}
B.~Andersson {\it et al.},
Phys.\ Rept. {\bf 97} (1983) 31.

\bibitem{JETSETandPYTHIA}
T.~Sj\"ostrand {\it et al.},
Comput. Phys. Commun. {\bf 82} (1994) 174
[hep-ph/9508391].

\bibitem{PYTHIA}
T.~Sj\"ostrand {\it et al.},
Comput. Phys. Commun. {\bf 135} (2001) 238
[hep-ph/0010017].


\bibitem{ALEPH}
S.~Schael {\it et al.}~[ALEPH~Collaboration],
\newblock  Phys. Lett. {\bf B606} (2005) 265.

\bibitem{profftune}
A.~Buckley {\it et al.},
Eur. Phys. J. {\bf C65} (2010) 331 [arXiv:0907.2973].


\bibitem{clustermodel1}
B.R. Webber,
Nucl. Phys. {\bf B238} (1984) 492;  \\
G. Marchesini and B.R. Webber, Nucl. Phys. {\bf B310} (1988) 461.


\bibitem{HERACLES}
A.~Kwiatkowski, H.~Spiesberger, and H.J. M\"ohring, {\sc Heracles}1.0, 
Comput. Phys. Commun. {\bf 69} (1992) 155.


\bibitem{GEANT}
R.~Brun {\it et al.},
\newblock {\sc Geant} 3, Technical Report CERN-DD/EE-84-1(1987).

\bibitem{myThesis}
A. Grebenyuk,  
{ \it ``Transverse Momentum of Charged Particles in low-$Q^2$ DIS at HERA"}, 
PhD thesis, Hamburg University, 2012,
(available at \\
\url{http://www-h1.desy.de/publications/theses_list.html}).

\bibitem{H1det1}
I.~Abt {\it et al.}~[H1~collaboration],
Nucl.\ Instrum.\ Meth. {\bf A386} (1997) 310.

\bibitem{H1det2}
I.~Abt {\it et al.}~[H1~collaboration],
\newblock  Nucl.\ Instrum.\ Meth. {\bf A386} (1997) 348.

\bibitem{SPACAL}
R.~D.~Appuhn {\it et al.}~[H1 SpaCal~Group],
\newblock  Nucl.\ Instrum.\ Meth. {\bf A386} (1997) 397.

\bibitem{CST34} 
D. Pitzl {\it et al.}, Nucl. Instrum. Meth. {\bf A454} (2000) 334 [hep-ex/0002044].

\bibitem{CST35}
B. List, Nucl. Instrum. Meth. {\bf A501} (2001) 49.

\bibitem{CIP}
J.~Becker {\it et al.}, 
Nucl. Instrum. Meth. {\bf A586} (2008) 190.

\bibitem{FTDJINST_FromKarin}
P.J. Laycock {\it et al.}, 
JINST 7 (2012) T8003 [arXiv:1206.4068].

\bibitem{BST37}
J. Kretzschmar, {\it A Precision Measurement of the Proton Structure Function $F_2$ with the H1
Experiment}, PhD thesis, Humboldt University, Berlin, 2008,
(available at \\
\url{http://www-h1.desy.de/publications/theses_list.html}).

\bibitem{FST36} 
I. Glushkov, {\it ``D* Meson Production in Deep Inelastic Electron-Proton Scatter-
ing with the Forward and Backward Silicon Trackers of the H1 Experiment at
HERA''}, PhD thesis, Humboldt University, Berlin, 2007, 
(available at \\
\url{http://www-h1.desy.de/publications/theses_list.html}).


\bibitem{test-beam-Spacal}
T.~Nicholls {\it et al.} [H1 SpaCal Group],
Nucl.\ Instrum.\ Meth. {\bf A374} (1996) 149.

\bibitem{Lar}
B. Andrieu {\it et al.} [H1 Calorimeter Group], Nucl. Instrum. Meth. {\bf A336} (1993) 460.

\bibitem{test-beam-Lar}
B.~Andrieu {\it et al.}~[H1 Calorimeter~Group], 
Nucl.\ Instrum.\ Meth. {\bf A336} (1993) 499.


\bibitem{eSigma}
U. Bassler and G. Bernardi, 
Nucl. Instrum. Meth. {\bf A426} (1999) 583 [hep-ex/9801017].


\bibitem{hadroo2}
M. Peez, { \it ``Search for Deviations from the Standard Model in High Transverse Energy
Processes at the Electron Proton Collider HERA''} (In French), PhD thesis, Lyon University, 2003,
(available at \\
\url{http://www-h1.desy.de/publications/theses_list.html}); \\  
B. Portheault, { \it ``First Measurement of Charged and Neutral Current Cross Sections with the
Polarised Positron Beam at HERA II and QCD-Electroweak Analyses''} (In French), PhD thesis,
Paris XI ORSAY University, 2005, (available at \\
\url{http://www-h1.desy.de/publications/theses_list.html}); \\
S.~Hellwig,  { \it ``Untersuchung der $D^\ast-\pi_{\rm slow}$ Double Tagging
  Methode in Charmanalysen''},
 Dipl. thesis, Hamburg University, 2004,  
(available at \\
\url{http://www-h1.desy.de/publications/theses_list.html}).


\bibitem{ElCalib}
F.D.~Aaron {\it et al.}~[H1~Collaboration],
Eur.Phys. J. {\bf C63} (2009) 625 [arXiv:0904.0929].

\bibitem{HFSCalib}
D. Salek,  { \it ``Measurement of the Longitudinal Proton Structure Function in Diffraction at the H1 Experiment and Prospects for Diffraction at LHC''}, PhD thesis, Charles Univ. Prague, DESY-THESIS-2011-013 (available at \url{http://www-h1.desy.de/publications/theses_list.html}).

\bibitem{HERAPDF10}
F.D.~Aaron {\it et al.}~[H1 and ZEUS Collaboration],
JHEP {\bf 1001} (2010) 109 [arXiv:0911.0884].

\bibitem{CTEQ66M}
P.M.~Nadolsky {\it et al.},
Phys. Rev. {\bf D78} (2008) 013004 [arXiv:0802.0007].

\bibitem{GRV98}
M.~Gl\"uck, E.~Reya and A.~Vogt,
Eur. Phys. J. {\bf C5} (1998) 461 [hep-ph/9806404].


\end{thebibliography}
\end{document}